\documentclass[12pt,a4paper]{article}%
\usepackage{cite}
\usepackage{calc}
\usepackage[pdftex]{color,graphicx}
\usepackage{fullpage,epsf, amssymb} 
\usepackage{mathrsfs, epstopdf, amsmath} 
\usepackage{hyperref}
\usepackage{multirow}
\usepackage{caption}
\usepackage{subcaption}
\usepackage{soul}
\usepackage{braket}
\usepackage{slashed}
\usepackage{appendix}
\usepackage{cancel}
\usepackage[parfill]{parskip}
\usepackage{amsfonts}
\usepackage[pdftex]{color,graphicx}
\graphicspath{{figs/}}
\numberwithin{equation}{section} 
\newcommand{\hhref}[1]{\href{http://arxiv.org/abs/#1}{arXiv:#1}}
\newcommand{\beq}{\begin{equation}}
\newcommand{\eeq}{\end{equation}}

\usepackage{tikz}
\usepackage{tkz-euclide}
\usetkzobj{all}
\usetikzlibrary{decorations.pathmorphing}	
\tikzset{
    v/.style={decorate, decoration={snake, segment length=3mm, amplitude=0.75mm}, draw},
    f/.style={draw=black, postaction={decorate},
        decoration={markings,mark=at position .55 with {\arrow[very thick]{latex}}}},
    fb/.style={draw=black, postaction={decorate},
        decoration={markings,mark=at position .55 with {\arrowreversed[very thick]{latex}}}},
    fnar/.style={draw=black},
    g/.style={decorate, draw=black,
        decoration={coil,amplitude=3pt, segment length=3.5pt}},
    s/.style={dashed,draw=black, postaction={decorate},
        decoration={markings,mark=at position .55 with {\arrow[very thick]{latex}}}},
    sb/.style={dashed,draw=black, postaction={decorate},
        decoration={markings,mark=at position .55 with {\arrowreversed[draw=black,very thick]{latex}}}},
    snar/.style={dashed,draw=black,line width =1.25pt},
}

\newcounter{qnumber}

\newcommand{\bea}{\begin{eqnarray}}
\newcommand{\eea}{\end{eqnarray}}
\newcommand{\eq}[1]{Eq.~(\ref{#1})}

\newcommand{\cLL}{c_{LL}}
\newcommand{\cRR}{c_{RR}}
\newcommand{\cLt}{c_{L}}
\newcommand{\cLb}{c_{L^{b}}}
\newcommand{\cRt}{c_{R}}
\newcommand{\cRb}{c_{R^{b}}}
\newcommand{\cV}{c_V}
\newcommand{\ct}{c_t}
\newcommand{\ch}{c_{3}}
\newcommand{\cLLh}{c_{LL}^{h}}
\newcommand{\cRRh}{c_{RR}^{h}}
\newcommand{\cLth}{c_{L}^h}
\newcommand{\cLbh}{c_{L^{b}}^{h}}
\newcommand{\cRth}{c_{R}^h}
\newcommand{\cRbh}{c_{R^{b}}^{h}}
\newcommand{\cth}{c_t^h}

\newcommand{\cunoL}{\bar c^{\,(1)}_{L}}
\newcommand{\ctreL}{\bar c^{\,(3)}_{L}}
\newcommand{\cL}{\bar c_{L}}
\newcommand{\cR}{\bar c_{R}}
\newcommand{\ctR}{\bar c_{R}}
\newcommand{\cbR}{\bar c^{\,b}_{R}}
\newcommand{\ctbR}{\bar c^{\,tb}_{R}}
\newcommand{\cu}{\bar c_{u}}
\newcommand{\cH}{\bar c_{H}}
\newcommand{\csei}{\bar c_{6}}

\newcommand{\GeV}{\,\mathrm{GeV}}
\newcommand{\TeV}{\,\mathrm{TeV}}

\begin{document}
\hfill{CERN-PH-TH-2015-265}

\hfill{DFPD-2015-TH-26}

\renewcommand{\thefootnote}{\fnsymbol{footnote}}
\color{black}
\vspace{1.5cm}
\begin{center}
{\LARGE \bf Strong {\bf \it tW} Scattering at the LHC}\\[5mm]
\bigskip\color{black}
{{\bf Jeff Asaf Dror,$^{a}$    Marco Farina,$^{a,b}$ Ennio Salvioni,$^{c}$ and Javi Serra$^{d,e}$  \footnote{Email: \url{ajd268@cornell.edu},~ \url{farina@physics.rutgers.edu},~\url{esalvioni@ucdavis.edu},~\url{javier.serra.mari@cern.ch}}}
} \\[5mm]
{\it  (a)  LEPP, Department of Physics, Cornell University,\\Newman Laboratory, Ithaca, NY 14853, USA}\\[3mm]
{\it  (b)  New High Energy Theory Center, Department of Physics, Rutgers University,\\136 Frelinghuisen Road, Piscataway, NJ 08854, USA}\\[3mm]
{\it  (c)  Department of Physics, University of California, Davis,\\One Shields Avenue, Davis, CA 95616, USA}\\[3mm]
{\it  (d)  Dipartimento di Fisica e Astronomia, Universit\`a di Padova and\\INFN, Sezione di Padova, Via Marzolo 8, 35131 Padova, Italy}\\[3mm]
{\it  (e)  Theoretical Physics Department, CERN, 1211 Geneva 23, Switzerland}\\[3mm]
\end{center}
\bigskip
\bigskip
\bigskip
\centerline{\bf Abstract}
\begin{quote}
Deviations of the top electroweak couplings from their Standard Model values imply that certain amplitudes for the scattering of third generation fermions and longitudinally polarized vector bosons or Higgses diverge quadratically with momenta. This high-energy growth is a genuine signal of models where the top quark is strongly coupled to the sector responsible for electroweak symmetry breaking. We propose to profit from the high energies accessible at the LHC to enhance the sensitivity to non-standard top-$Z$ couplings, which are currently very weakly constrained. To demonstrate the effectiveness of the approach, we perform a detailed analysis of $tW \to tW$ scattering, which can be probed at the LHC via $pp\to t\bar{t}Wj$. By recasting a CMS analysis at $8 \TeV$, we derive the strongest direct bounds to date on the $Ztt$ couplings. We also design a dedicated search at $13 \TeV$ that exploits the distinctive features of the $t\bar{t}Wj$ signal. Finally, we present other scattering processes in the same class that could provide further tests of the top-Higgs sector.
\end{quote}

\renewcommand{\thefootnote}{\arabic{footnote}}\setcounter{footnote}{0}

\enlargethispage{1cm}

\pagestyle{empty}

\newpage

\section{Introduction} \label{sec:}
\setcounter{page}{1}
\pagestyle{plain}
The large center of mass energies accessible at the LHC make it the optimal machine to explore the electroweak scale. This has already been confirmed by the discovery of the Higgs boson \cite{ATLASdiscovery,CMSdiscovery}, which represents the main achievement of Run-1 and a major step forward for particle physics. Another important example of the power of the LHC is the large rate for production of the top quark, the particle in the Standard Model (SM) with the largest coupling to the Higgs field. However, our knowledge of the properties of both the Higgs and the top is still relatively poor. Since these two particles play a central role in theories beyond the SM (BSM) that provide a deeper understanding of electroweak symmetry breaking (EWSB), the program of Higgs and top coupling measurements is one of the priorities of LHC Run-2. The importance of this task is reinforced by the thus far lack of evidence for direct production of BSM particles, which may suggest that probes of the Higgs and top sectors are our best opportunity to gain new insights into the mechanism of EWSB. 

The complicated hadronic environment at the LHC, however, does not facilitate the desired experimental precision. For example, experimental tests of the $Ztt$ and $htt$ couplings are very challenging: the conventional strategy consists in measuring the cross section for $t\bar{t}Z$ and $t\bar{t}h$ production, respectively. These processes have a relatively high mass threshold and thus suppressed production rates at the LHC. This leads to very loose constraints on the top couplings, currently well above the SM expectations. On the other hand, projections indicate that the $htt$ coupling could be measured with $15\%$ accuracy by the end of Run-2 \cite{SnowmassHiggs}, whereas the expected precision on $Ztt$ is worse, and deviations as large as $50$-$100\%$ will not be excluded \cite{ttZoriginal1,LowETAL,RS}.

One is then prompted to ask if there exists another avenue to probe the properties of the Higgs and the top. 
An answer has been given already for the couplings of the Higgs boson: if the Higgs couplings to the electroweak gauge bosons depart from the SM predictions, the amplitudes for the scattering of the longitudinally polarized $V = W, Z$ and Higgs $h$ undergo a rapid growth with momentum above the weak scale $v \simeq 246 \GeV$. The prime example is $VV \to VV$ scattering, which grows with momenta as $p^2/v^2$ whenever the $hVV$ coupling deviates from the SM\cite{WWscattering}, while the process $VV \to hh$ provides complementary information \cite{strongh}. Such growth with energy is a distinctive feature of models where the Higgs emerges as a pseudo-Goldstone boson from a strongly-coupled sector \cite{silh,reviewUS,reviewEU}. In this class of theories, the high-energy enhancement can be accessed without directly producing the BSM resonances, which are strongly coupled to the Higgs but heavy. In complete analogy, the electroweak couplings of the top could be probed in the high-energy scattering of third generation fermions and longitudinal gauge bosons or Higgses. A growth with energy of the associated amplitudes would constitute a genuine signal of the strong coupling of the top to the BSM sector \cite{LillieETAL,compot,KumarTaitVM}. It was observed a long time ago \cite{AppelquistChanowitz} that a deviation from the SM in the $h \psi \psi$ coupling (with $\psi$ a SM fermion) leads to a growth proportional to $m_\psi p/v^2$ of scattering amplitudes such as $\psi \bar{\psi} \to VV$, and this observation was recently exploited in Refs.~\cite{Biswasth1,Farinath} to constrain the $htt$ coupling at the LHC, via the scattering $bW\to th$ in the $pp\to thj$ process \cite{Maltonith}.

In this paper we perform a general analysis of the scattering of tops (and bottoms) with the longitudinal $W, Z$ or the Higgs. We point out that in the presence of deviations in the $Ztt$ couplings, certain amplitudes grow like $p^2/v^2$ (rather than $m_t p/v^2$), leading to an enhanced sensitivity at the LHC. The $tW \to tW$ amplitude is singled out as the most promising one, because deviations in either the $Zt_L t_L$ or $Zt_R t_R$ couplings lead to the strong high-energy behavior. Furthermore, the corresponding LHC process is $pp\to t\bar{t}Wj$, which gives a clean same-sign leptons signature. We perform a detailed analysis of this signal, exploiting the information contained in the CMS $8 \TeV$ search for $t\bar{t}W$ of Ref.~\cite{CMSttV}, and show that it gives stronger constraints than the conventional strategy relying on $pp\to t\bar{t}Z$. Motivated by the effectiveness of our approach at $8 \TeV$, we then design a specific search for Run-2 at $13$ TeV, which we hope will help in refining the physics analyses of the experimental collaborations. 

We also interpret our analysis in terms of non-standard top couplings arising from dimension-6 (dim-$6$) operators added to the SM Lagrangian, and show that competitive bounds are obtained in this case too. In this framework, correlations arise between the couplings of the top to the $Z$ and to the $W$. Moreover, deviations in these couplings imply a $p^2/v^2$ growth not only of the $tV$ scattering amplitudes, but also of those involving the Higgs, such as $bW\to th$ \cite{windows}. Thus, the interest of our approach does not end here: we discuss several other amplitudes that we believe to be promising in probing the top electroweak couplings, and that warrant further work to assess the expected sensitivity at the LHC.

Our paper is organized as follows. In Sec.~\ref{sec:CouplingsandEFT} we introduce our parameterization of couplings in the top-Higgs sector, discuss the current experimental constraints and outline the generic aspects of the scattering of third generation fermions with electroweak vector bosons and Higgses. Section~\ref{sec:tWtW} contains the discussion of the $tW\to tW$ scattering and the associated LHC process $pp\to t\bar{t}Wj$, as well as the main results of our paper. Section~\ref{sec:analysis} contains the technical details of our collider analysis of $t\bar{t}Wj$, as well as the description of the method we use to obtain constraints on the top-$Z$ interactions. This section can be omitted by the reader interested only in the results, who can move on to Sec.~\ref{sec:OtherProcesses}, where we discuss other scattering processes that may provide additional information on the top-Higgs sector. Finally, Sec.~\ref{sec:conclusions} contains our conclusions. Three Appendices complete the paper: App.~\ref{App:chiL} presents the electroweak chiral Lagrangian for the top sector, App.~\ref{App:ttZ} summarizes the current and projected constraints on top-$Z$ couplings obtained from $pp\to t\bar{t}Z$, and App.~\ref{App:FakeLeptons} details the procedure we adopt to simulate `fake' leptons, which constitute one of the main backgrounds to our $t\bar{t}Wj$ signal.
\section{Parameterization of top and Higgs couplings}\label{sec:CouplingsandEFT}
In this section we introduce the general parameterization of the couplings relevant for the scattering of top quarks with the electroweak vector bosons $W$ and $Z$ and with Higgs boson $h$. The interactions of the top (and bottom) are encoded in the phenomenological Lagrangian
\bea
\label{Ltphen}
\mathcal{L}_{t} \!\!\!&=&\!\!\! Z_\mu \, \bar t \gamma^\mu \big[ \cLt(h) \, g_{Zt_Lt_L} P_L + \cRt(h) \, g_{Zt_Rt_R} P_R \big]t  + Z_\mu \, \bar b \gamma^\mu \big[ \cLb(h) \, g_{Zb_Lb_L} P_L + \cRb(h) \, g_{Zb_Rb_R} P_R \big]b \nonumber \\
&&\!\!\! + \ g_{Wt_Lb_L} W^+_\mu \bar t \gamma^\mu \big[ \cLL(h) P_L + \cRR(h) P_R \big]b + \mathrm{h.c.} - \ct(h) \frac{m_t}{v} h \, \bar t t \ , 
\eea
where $P_{L,R}$ are the left ($L$) and right ($R$) chiral projectors, $g_{Wt_Lb_L } = g/\sqrt{2}$, $g_{Zf_Rf_R} = -(g s_w^2/c_w) Q_f$, $g_{Zf_Lf_L} = (g/c_w) (T_{L,f}^3-Q_f s_w^2)$ are the SM gauge couplings, and $v \simeq 246 \GeV$. 
We have defined the coefficients above as linear functions of $h$,
\beq
\label{chat}
c_i(h) \equiv c_i + 2 c_i^h \frac{h}{v} \ ,
\eeq
($i = \{L, R, L^b, R^b, LL, RR, t\}$), such that they also encode BSM couplings of the Higgs. 
We will also describe the $hVV$ and Higgs cubic couplings %
with the Lagrangian
\beq
\label{Lhphen}
\mathcal{L}_{h} = \cV \frac{m_W^2}{v} h \left( 2 W^+_\mu W^{-\mu} + \frac{1}{c_w^2} Z_\mu Z^\mu \right) 
- \ch \frac{m_h^2}{2v} h^3 \ .
\eeq
The coefficients $c_i$, $c_i^h$, $\cV$ and $\ch$ parameterize the relevant couplings of the third generation fermions, $W$, $Z$, and $h$. 
The SM Lagrangian is reproduced for
\beq
\label{SMc}
\cLt = \cRt = \cLb = \cRb = \cLL = \ct = \cV = \ch = 1 \ , \quad \cRR = 0 \ , \quad c_i^h = 0\,.
\eeq
We now wish to comment on the rationale behind our parameterization. As explained in App.~\ref{App:chiL}, the phenomenological Lagrangian in Eqs.~(\ref{Ltphen}) and (\ref{Lhphen}) can be regarded as the unitary gauge version of the leading set of operators, in an expansion in derivatives, of the electroweak chiral Lagrangian \cite{PecceiZhang,MalkawiYuan} (for recent thorough discussions of the electroweak chiral Lagrangian, see Refs.~\cite{strongh,nonlinear}). We are neglecting, for instance, BSM chirality-flipping interactions of the fermions with $W$ and $Z$, which arise at the next order in the derivative expansion. Denoting by $\Lambda$ the mass scale of the new physics resonances, such interactions are generically suppressed by $p/\Lambda$ with respect to the ones we consider here, with $p$ characterizing the momenta of the process. Due to the chirality flip, they are further suppressed by $y_t/g_*$, with $g_*$ a generic BSM coupling satisfying $g_* \leqslant 4 \pi$. A notable class of chirality-flipping interactions are dipole-type operators, whose schematic structure is, for example, $\sim \bar{t}_L \sigma^{\mu\nu}t_R Z_\mu p_\nu$. In addition to the previous considerations, dipole operators are not generated at tree level if the transverse SM gauge fields are external to the BSM sector and coupled to it through weak gauging of the corresponding symmetries, as we assume.\footnote{Besides, constraints on top dipole moments, either direct from top decay and single top production measurements \cite{dipole1}, or indirect from the experimental limits on $b \to s$ transitions \cite{dipole2}, are already significant.} We also set the triple gauge interactions to their SM values. We choose to do so because in theories where the SM gauge bosons are weakly coupled to the BSM sector, generic deformations of the triple gauge interactions yield small effects in the processes we are interested in \cite{windows}.\footnote{Additionally, current bounds on these couplings are already below $10\%$, and improved sensitivities from diboson production measurements are expected at the $13 \TeV$ LHC run \cite{tgcs}.} 
Finally, we will also be neglecting the small effects due to the bottom Yukawa coupling.

Theories where the typical scale of the BSM sector can be decoupled from the electroweak scale, $\Lambda/g_* \gg v$, admit a further expansion in the Higgs doublet field $H$. 
In such a case BSM effects from heavy resonances can be parameterized by operators of dimension larger than four built out of the SM fields. 
We are particularly interested in the dim-6 operators\cite{silh,compot}
\bea
\Delta \mathcal{L}_{t} \!\!\!&=&\!\!\! 
\frac{i \cunoL}{v^{2}} H^{\dagger} \overleftrightarrow{D_{\mu}} H \bar{q}_{L}\gamma^{\mu}q_{L} 
+ \frac{i \ctreL}{v^{2}} H^{\dagger} \sigma^{a} \overleftrightarrow{D_{\mu}} H \bar{q}_{L}\gamma^{\mu}\sigma^{a}q_{L} \nonumber \\
&&\!\!\! + \ \frac{i \ctR}{v^{2}} H^{\dagger} \overleftrightarrow{D_{\mu}} H \bar{t}_{R}\gamma^{\mu}t_{R} + \frac{i \cbR}{v^{2}} H^{\dagger} \overleftrightarrow{D_{\mu}} H \bar{b}_{R}\gamma^{\mu}b_{R}+ \left(\frac{i \ctbR}{v^{2}} \tilde H^{\dagger} \overleftrightarrow{D_{\mu}} H \bar{t}_{R}\gamma^{\mu}b_{R} + \mathrm{h.c.}\right) \nonumber \\
&&\!\!\! + \ \frac{\cu y_t}{v^2} H^\dagger H \bar q_L \tilde{H} t_R + \mathrm{h.c.} \ , 
\label{Lsilh}
\eea
where $\tilde H = i\sigma_2 H^*$ and we defined $H^{\dagger} \overleftrightarrow{D_{\mu}} H \equiv H^{\dagger} (D_{\mu} H) - (D_\mu H)^{\dagger} H$, etc..
These operators modify the couplings of the top (and bottom) to the $W$, $Z$, and $h$ with respect to the SM, such that
\beq \label{Lcoup}
c_{L}-1 = \cLth = \frac{\ctreL-\cunoL}{1-\frac{4}{3}s_w^2}
 \ , \quad 
c_{L^{b}} -1 = \cLbh = \frac{\cunoL+\ctreL}{1-\frac{2}{3}s_w^2}
 \ , \quad 
c_{LL} - 1 = \cLLh = \ctreL
 \ , 
\eeq
\beq \label{Rcoup}
c_{R} -1 = \cRth = \frac{\ctR}{\frac{4}{3}s_w^2}
 \ , \quad 
c_{R^{b}} - 1 = \cRbh = -\frac{\cbR}{\frac{2}{3}s_w^2}
 \ , \quad  
\cRR = \cRRh = \ctbR
 \ , 
\eeq
\beq
c_t -1 = \tfrac{4}{3} \cth = - \cu\,.
\label{deltag}
\eeq
Notice that at the dim-6 level none of the $c_i^h$ coefficients is independent from the $c_i$'s. 
Furthermore, while each of the $R$-handed couplings in Eq.~\eqref{Rcoup} is affected by an independent dim-6 operator, the deviations in the $L$-handed ones, Eq.~\eqref{Lcoup}, are partially correlated. 
This is due to a remnant custodial symmetry of the dim-6 Lagrangian, which is broken by dim-8 operators \cite{bsmprimaries}, or absent altogether in the electroweak chiral Lagrangian, see App.~\ref{App:chiL}.
The dim-6 operators giving rise to non-standard contributions to the terms in \eq{Lhphen} can be written as
\beq
\Delta \mathcal{L}_{h} = \frac{\cH}{2 v^2} (\partial_\mu |H|^2 )^2 - \frac{\csei \lambda}{v^2} |H|^6 \ ,
\label{Lsilh2}
\eeq
with $\cV - 1 = -\cH/2$ and $\ch -1 = \csei - 3 \cH/2$. 
The operator $\cH$ also contributes to $c_t$ by an amount $- \cH/2$. 
The set of dim-6 operators in Eqs.~(\ref{Lsilh}) and (\ref{Lsilh2}) encode the leading non-standard effects in theories where both the Higgs and either the $L$- or $R$-handed top are strongly coupled to a BSM sector whose generic coupling strength is $g_* > g_{\mathrm{SM}}$, with $g_{\mathrm{SM}}$ the weak gauge or top Yukawa couplings. 
In such scenarios the corresponding $\bar c$ coefficients can be as large as
\beq 
\bar c \lesssim \frac{g_*^2 v^2}{\Lambda^2} \equiv \xi
\label{coeff}
\eeq
with $g_* \leqslant 4 \pi$, barring $O(1)$ factors.
Particularly relevant examples of such a situation are composite Higgs models with top partial compositeness \cite{reviewUS,reviewEU}.%
\footnote{In those models the Higgs field arises as a Nambu-Goldstone boson, and the parameter $\xi$ defined in \eq{coeff} is identified with $v^2/f^2$, where $f$ is the Higgs decay constant.}
In such models the need to reproduce the large top Yukawa coupling forces one or both of the top chiral states to couple strongly to the composite sector. 
We would also like to stress that when $g_* \gg g_{\mathrm{SM}}$, the relative importance of probing non-standard top couplings versus direct searches for BSM resonances increases, given that larger values of the resonance mass $\Lambda$ can be considered.%
\footnote{This is of special relevance, for instance, in composite Twin Higgs models, where the composite resonances, despite being heavy, remain strongly coupled to the Higgs and the top \cite{Geller:2014kta,compotwin,Low:2015nqa}.}

Out of the BSM effects introduced above, in this work we will mostly focus on the couplings of the top to the $Z$, $\cLt$ and $\cRt$, not only because of their impact on top scattering processes, but also because they are very weakly constrained by direct measurements. 
Up to date, the only bound comes from the analysis of $t\bar{t}Z$ production at the $7$ TeV LHC \cite{RS}, from which $O(1)$ deviations in $c_L$ or $c_R$ cannot be excluded.  
In contrast, other BSM effects are already subject to significant constraints.
The most stringent one is on the $Z b b$ coupling: LEP1 measurements directly constrain $c_{L^{b}}$ at the per-mille level, while the bound on $c_{R^{b}}$ is at a few per-cent \cite{Zbb}. Due to the former constraint, BSM sectors are typically assumed to couple to $q_L$ such that a custodial $P_{LR}$ symmetry is preserved \cite{custodialparity}, yielding $\cLb = 1$ to leading approximation. 
In terms of dim-6 operators, this implies $\cunoL = - \ctreL$. On the other hand, direct bounds on the $Wtb$ coupling coefficients $c_{LL}$ and $c_{RR}$ from single top production \cite{TaitYuan} and $W$ helicity fraction measurements are around $10\%$ \cite{dipole1,Buckley:2015nca}. 
 Notice that in terms of dim-6 operators, the combined constraints on the $Z b_L b_L$ and $Wt_L b_L$ couplings, which bound both $\cunoL$ and $\ctreL$, imply BSM effects of at most $\sim 10\%$ on the $Zt_L t_L$ coupling. However, it should be kept in mind that the experimental status is not yet such as to fully motivate the hypothesis of a large new physics scale $\Lambda$ compared to the electroweak scale, at least for what regards direct probes of the top sector.

We now turn to the discussion of the indirect bounds. The $L$- and $R$-handed top couplings to the $W$ and $Z$ are indirectly probed by electroweak precision data, via top loop contributions to the $\widehat S$ and $\widehat T$ parameters as well as to the $Z b_L b_L$ coupling, all of which have been measured with per-mille accuracy. 
The contribution of $\cunoL$, $\ctreL$, $\ctR$ to the renormalization group running of the dim-6 operators associated to the aforementioned observables can be consistently computed within the effective theory \cite{compot,windows}. 
For instance, assuming $\ctreL = -\cunoL$ at the scale $\Lambda$, the $\widehat T$-parameter is renormalized by $\Delta \widehat T = N_c y_t^2 (\cunoL-\ctR) \log(\Lambda/\mu) /(4 \pi^2)$, and similar log-divergent terms are generated for $\widehat S$ and $Z b_L b_L$.
Taken at face value, this set of contributions imply the bounds $\cunoL, \ctR \lesssim 5 \%$ \cite{topewpd}. This is analogous to the indirect bound set on $\cH$ from log-divergent Higgs loop contributions to $\widehat S$ and $\widehat T$ \cite{sthiggs}, which nevertheless does not undermine the relevance of a direct measurement of the $hVV$ ($V = W, Z$) coupling at the LHC. The same logic should apply to direct measurements of the top-$Z$ couplings, even more so after taking into account that, in the cases of interest in this work, $\widehat S$, $\widehat T$ and $Zb_L b_L$ are dominated by incalculable ultraviolet (UV) contributions: Since it is not protected by any symmetry, $\widehat S$ generically receives UV contributions at tree level. On the other hand, even though $\widehat T$ and $Zb_L b_L$ can be UV protected if the BSM sector is custodial and $P_{LR}$ symmetric, contributions to $\widehat T$ from top loops with two insertions of $\cunoL$ or $\ctR$ are actually quadratically divergent and dominant whenever these coefficients are large. The situation is similar for loop contributions to $Zb_L b_L$ from one insertion of $\cunoL = - \ctreL$ and another of four-fermion operators \cite{compot}.\footnote{The four-fermion operators are irrelevant for the scattering processes we study in this work, but nevertheless large in the same type of BSM scenarios.} Finally, we briefly mention bounds from flavor observables. The $\bar{c}_R^{\,tb}$ coefficient contributes at one loop to the $b \to s \gamma$ decay rate, with an amplitude enhanced by $m_t/m_b$, and is thus constrained at the per-mille level. In addition, $Z$-mediated penguin contributions to rare $B$ and $K$ meson decays lead to constraints on $\cunoL$, $\ctreL$ and $\ctR$ \cite{tZflavor}, which are at the same level of those from electroweak precision data. All these bounds, however, strongly depend on the assumed underlying flavor structure. In conclusion, currently little can be said with confidence about the couplings of the top to the $Z$, which motivates the new approach for probing them presented in this work.

As far as the couplings of the Higgs boson are concerned, constraints are still relatively mild. Global fits to inclusive signal strengths give $\cV \lesssim 20 \%$ \cite{higgscombo}, whereas searches for the $t\bar{t}h$ signal still allow $O(1)$ deviations in $c_t$ \cite{ATLAS_ttH_bb,ATLAS_ttH_gamma,ATLAS_ttH_multilepton,CMS_ttH}. On the other hand, no experimental constraint currently exists on the Higgs cubic coupling $c_3$, nor on the $c_{i}^{h}$ defined in Eqs.~(\ref{Ltphen}, \ref{chat}).

The strength of the constraints discussed above relies on the relative precision of the experimental measurements compared to the BSM effects, which are of size $\xi$ or smaller, see Eq.~\eqref{coeff}. In particular, the large uncertainty that affects the LHC measurement of the $t\bar{t}Z$ production cross section is behind the weakness of the direct bounds on modified top-$Z$ couplings. However, there is another avenue for constraining non-standard top interactions, which relies on the large center of mass energies that can be reached at the LHC: departures from the SM prediction of certain top couplings imply that some scattering amplitudes will diverge with the momenta of the process. An analogy can be drawn with $VV \to VV$ scattering, where non-SM values of $\cV$ lead to a growth of the amplitude with energy. In our case, the scatterings of interest are $tV \to tV$ and its crossings. Both in $VV$ and $tV$ scattering, the amplitudes that grow the most with energy involve the longitudinal polarizations of the $W^\pm$ and the $Z$. For $tV$ scattering this can be clearly seen by inspecting the interactions of the top in a gauge where the Nambu-Goldstone bosons eaten by the $W$ and $Z$, which we label $\chi_a$ ($a=1,2,3$), appear explicitly in the Lagrangian, see Eqs.~(\ref{Lchitop}) and (\ref{devsigmagaugeless}) in App.~\ref{App:chiL}. For non-SM values of $\cLt$, $\cRt$, etc., four-point contact interactions of the form $\epsilon_{abc} \chi_b \partial_\mu \chi_c (\bar \psi \gamma^\mu \psi)_a / v^2$, with $\psi = \{t, b\}$, are generated, implying a $p^2/v^2$ growth of the amplitudes $\psi \chi \to \psi \chi$. Notice that the symmetry structure of the interaction is such as to include, for example, $t W^\pm \to t W^\pm$, but not $t Z \to t Z$. Likewise, certain scattering amplitudes involving the Higgs, such as $b W^+ \to t h$, also display the same divergent behavior at high energies. This follows from the interactions $h \partial_\mu \chi_a (\bar \psi \gamma^\mu \psi)_a / v^2$, also shown in App.~\ref{App:chiL}. The relation between the $tV$ and $th$ scattering amplitudes is also obvious when interpreted in terms of dim-6 operators, given the relations in Eqs.~(\ref{Lcoup}--\ref{deltag}). The $p^2/v^2$ growth should be contrasted with the $m_t p/v^2$ growth that arises if the Higgs couplings $c_t$ or $c_V$ deviate from the SM \cite{AppelquistChanowitz} (see also \cite{Grinstein:2013fia}), whereas no enhancement with energy is generated by deviations in the Higgs cubic coupling $c_3$. Thus, Higgs coupling modifications only give subleading effects in the high-energy scattering processes we are interested in.

To summarize, in certain two to two scattering processes the sensitivity to non-standard top-$Z$ couplings is enhanced at high energies, possibly overcoming the limited experimental precision. The enhancement scales as $\bar c \, p^2/v^2 \sim g_*^2 p^2/\Lambda^2$, which can be much larger than one in models where $g_* \gg 1$, without being in conflict with the effective field theory expansion, that is $p^2 < \Lambda^2$. This approach then takes advantage of the high scattering energies accessible at the LHC. We explicitly demonstrate its effectiveness in the next section, focusing on $tW \to tW$.

\section{$tW\to tW$ scattering as case study} \label{sec:tWtW}
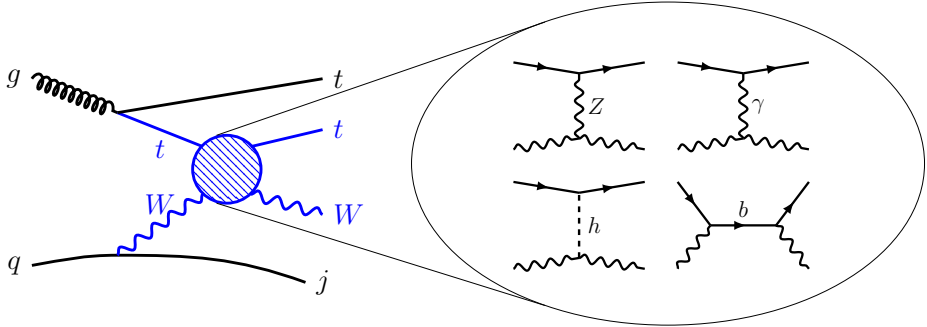
\begin{figure} 
\begin{center} 
\definecolor{myred}{rgb}{0, 0, 1}
\scalebox{0.9}{
 \begin{tikzpicture}
\begin{scope}[line width=1.2pt]  
\draw[g] (0,3) -- (1.25,2.5) node[pos=0,left] {$g$}; 
\draw[fnar,myred] (1.25,2.5) -- (2.5,2) node[midway,below] {$t$}; 
\draw[](0,0.25) node[left] {$ q $} to [out=10,in=180] (1.25,0.4) to [out=0,in=160] (4,0) node[right] {$ j $};
\draw[v,myred] (1.25,0.4) -- (2.5,1.3) node[pos=0.5,left,above] {$W  $}; 
\draw[myred,pattern=north west lines,pattern color=myred,very thick] (2.85,1.66667) circle (0.5);
\draw[fnar,myred] (3.2,2) -- (4.25,2.25) node[right] {$t$}; 
\draw[fnar] (1.25,2.5) -- (4.25,3) node[right] {$t $}; 
\draw[v,myred] (3.2,1.3) -- (4.25,1) node[right] {$W  $}; 
\end{scope}
\draw[] (2.7,1.6667+0.5) -- (7.5,3.835);
\draw[] (2.7,1.6667-0.5) -- (7.5,-0.34);
\draw[] (9.3,1.75) ellipse (3.75 and 2.38);

\scalebox{0.8}{\begin{scope}[line width = 1.2pt,shift={(0.5,0.25)}] 
\pgfmathsetmacro{\ymin}{-0.3}
\pgfmathsetmacro{\ymax}{1.3}
\pgfmathsetmacro{\xmin}{0.3}
\pgfmathsetmacro{\xmax}{2.7}
\begin{scope}[shift={(8,2.5)}]
  \draw[v] (\xmin,\ymin) 
  -- (1.5,\ymin+0.2);
\draw[f] (\xmin,\ymax) 
-- (1.5,\ymax-0.2);
\draw[v](1.5,\ymin+0.2) -- (1.5,\ymax-0.2) node[midway,right] {$ Z$};
\draw[v] (1.5,\ymin+0.2) -- (\xmax,\ymin); 
\draw[f] (1.5,\ymax-0.2) -- (\xmax,\ymax);
\end{scope}
\begin{scope}[shift={(11,2.5)}]
  \draw[v] (\xmin,\ymin) 
  -- (1.5,\ymin+0.2);
\draw[f] (\xmin,\ymax) 
-- (1.5,\ymax-0.2);
\draw[v](1.5,\ymin+0.2) -- (1.5,\ymax-0.2) node[midway,right] {$ \gamma $};
\draw[v] (1.5,\ymin+0.2) -- (\xmax,\ymin); 
\draw[f] (1.5,\ymax-0.2) -- (\xmax,\ymax);
\end{scope}
\begin{scope}[shift={(8,0.3)}]
  \draw[v] (\xmin,\ymin) 
  -- (1.5,\ymin+0.2);
\draw[f] (\xmin,\ymax) 
-- (1.5,\ymax-0.2);
\draw[snar](1.5,\ymin+0.2) -- (1.5,\ymax-0.2) node[midway,right] {$ h$};
\draw[v] (1.5,\ymin+0.2) -- (\xmax,\ymin); 
\draw[f] (1.5,\ymax-0.2) -- (\xmax,\ymax);
\end{scope}
\begin{scope}[shift={(11,0.3)}]
\def\xi{{\xmin+(\xmax-\xmin)/4}};
\def\xf{{\xmax-((\xmax-\xmin)/4)}};
\def\ymid{{\ymin+(\ymax-\ymin)/2}};

  \draw[v] (\xmin,\ymin) 
  -- (\xi,\ymid);
\draw[f] (\xmin,\ymax) 
-- (\xi,\ymid);
\draw[f] (\xi,\ymid) -- (\xf,\ymid) node[midway,above] {$ b $};
\draw[v] (\xf,\ymid) -- (\xmax,\ymin); 
\draw[f] (\xf,\ymid) -- (\xmax,\ymax);
\end{scope}
\end{scope}}
\end{tikzpicture}
}
\end{center} 
\caption{$t W \rightarrow t W $ scattering at the LHC. For definiteness, in the inset we show the diagrams corresponding to $tW^{-}\to tW^{-}$.}
\label{fig:FeynmanDiagrams}
\end{figure}

Our goal is to study the scattering amplitudes involving tops (and/or bottoms) and $W, Z$ or $h$ that increase at high energies, and to exploit this growth to probe top-$Z$ interactions. 
After examining all the possible combinations, we focus on the process $tW \to tW$. Our motivation for this choice is threefold:
\begin{enumerate}
\item The amplitude for $tW\to tW$ scattering grows with the square of the energy if either the $Zt_L t_L$ or the $Zt_R t_R$ couplings deviate from their SM values.
\item The corresponding collider process, $pp\to t\bar{t}Wj$, gives rise to same-sign leptons (SSL), an extremely rare final state in the SM. This process arises at $O(g_s g_w^3)$ in the gauge couplings, where $g_s$ denotes the strong coupling and $g_w$ any electroweak coupling, as shown in Fig.~\ref{fig:FeynmanDiagrams}. 
\item The main irreducible background, $pp\to t \bar{t}W + \mathrm{jets}$ at $O(g_s^{2+n}g_w)$ with $n\geq 0$ the number of jets, is insensitive to the details of the top sector, because the $W$ is radiated off a light quark.   
\end{enumerate}   
%
%

The amplitude for two to two scattering processes of the type $\psi_1 + \phi_1 \to \psi_2 + \phi_2$, where $\psi_{1,2} = \{t,b\}$ and $\phi_{1,2} = \{\chi^\pm \equiv (\chi_1 \mp i \chi_2)/\sqrt{2}$, $\chi_3,h\}$ are the longitudinal $W^{\pm}, Z$ or $h$, is most conveniently expressed in the basis of chirality eigenstate spinors. Retaining only terms that grow with energy, we find
\begin{equation} \label{GeneralAmplitude}
\begin{pmatrix} \mathcal{M}_{LL} & \mathcal{M}_{RL} \\ \mathcal{M}_{LR} & \mathcal{M}_{RR} \end{pmatrix} =  \frac{\kappa\, g^2}{2m_W^2} \begin{pmatrix} e^{i\varphi}\sqrt{\hat{s}(\hat{s}+\hat{t})}A_{LL} & m_{t} \sqrt{-\hat{t}} \,A_{RL} \\
-e^{i\varphi}m_t \sqrt{-\hat{t}} \, A_{LR} & \sqrt{\hat{s}(\hat{s}+\hat{t})} A_{RR} \end{pmatrix}  \ ,
\end{equation}
where $\kappa$ and $A_{ij}$ ($i,j = L,R$, with $i$ indicating the chirality of $\psi_1$ and $j$ the chirality of $\psi_2$) are process-dependent coefficients.\footnote{We take initial state momenta as ingoing, and final state momenta as outgoing. The Mandelstam variables are defined as $\hat{s}=(p_{\psi_1} + p_{\phi_1})^2$ and $\hat{t}=(p_{\phi_1} - p_{\phi_2})^2$, and $\varphi$ is the azimuthal angle around the $z$ axis, defined by the direction of motion of $\phi_1$.}
 In particular, the $A_{ij}$ encode the dependence on the anomalous couplings: $A_{LL}$ and $A_{RR}$ control the leading amplitudes, which grow as $\hat{s}$, whereas $A_{LR}$ and $A_{RL}$ control the subleading pieces, growing as $\sqrt{\hat{s}}$. All the $A_{ij}$ vanish in the SM, where the amplitude must tend to a constant limit at large $\sqrt{\hat{s}}$. For $tW^{-}\to tW^{-}$ scattering we have $\kappa = 1$ and
\begin{align}
A_{LL}&\,=\, - \cLL^2 + \cLt - \tfrac{4}{3} s_w^2 (\cLt-1)\,, \nonumber \\
A_{RR}&\,=\, - \cRR^2 - \tfrac{4}{3} s_w^2 (\cRt -1 )\,, \nonumber \\
A_{LR}&\,=\, A_{RL} = \tfrac{1}{2} \left[ (\cLt-\ct \cV ) - \tfrac{4}{3} s_w^2 (\cLt + \cRt-2) \right] . \label{tW-tW-}
\end{align}
For $t W^{+}\to t W^{+}$, we find again that $\kappa = 1$, $A_{LL}$ and $A_{RR}$ are identical to those in Eq.~\eqref{tW-tW-}, whereas the subleading pieces read
\begin{align}
A_{LR}\,=\,  A_{RL} = \cLL^2 + \cRR^2 - \tfrac{1}{2} \left[ (\cLt+\ct \cV ) - \tfrac{4}{3} s_w^2 (\cLt + \cRt-2) \right].
\end{align}
We see that whenever $c_L(c_R)\neq 0$ (and barring accidental cancellations), the $LL(RR)$ amplitude grows like $\hat{s}$. This has to be contrasted with the weaker growth like $\sqrt{\hat{s}}$ caused by deviations in the Higgs couplings $c_V$ or $c_t$. Because their effect is subleading, in our analysis of $tW$ scattering we will set $c_V = c_t = 1$, and focus exclusively on modifications of top-$Z$ interactions. For the latter we will consider two different theoretical interpretations. The first one targets the $Zt_L t_L$ and $Z t_R t_R$ couplings, by taking $\Delta_{L,R}\equiv c_{L,R}-1 \neq 0$ in Eq.~\eqref{deltag}, whereas all other coefficients are set to their SM values. 
Under this assumption, the leading terms in the amplitude read
\begin{equation} \label{ampcoup}
A _{LL} =  \left(1  - \tfrac{4}{3} s_w ^2 \right)  \Delta _L\,,  \qquad \qquad  A _{RR} = - \tfrac{4}{3} s _w ^2 \Delta _R\,.
\end{equation} 
We note that the sensitivity to $\Delta _R$ is lower than to $\Delta _L$ due to the $s_w ^2$ suppression of $A_{RR}$. 
In addition, we present results in the framework of higher-dimensional operators (HDO), where the deviations in the top-$Z$ couplings are correlated with those in other interactions of the third generation fermions. As discussed in Sec.~\ref{sec:CouplingsandEFT}, the per-mille constraint on the $Zb_L b_L$ vertex forces us to assume $\ctreL = - \cunoL$. We thus take $\cL \equiv \cunoL = - \ctreL$ and $\cR$ as BSM parameters, whereas all the other $\bar{c}_i$ coefficients in Eqs.~(\ref{Lsilh}) and (\ref{Lsilh2}) are set to zero. Notice that under these assumptions, $\bar{c}_L$ also modifies the $Wt_L b_L$ vertex, which contributes to $tW\to tW$ scattering via the $b$-exchange diagram in Fig.~\ref{fig:FeynmanDiagrams}. The leading amplitudes read 
\begin{equation} \label{ampEFT} 
A _{LL} = -\cL^{\,2}\,, \qquad\qquad  A _{RR} = - \cR\,.
\end{equation} 
We note that in $A_{LL}$ the term linear in $\cL$ vanishes. This can be traced back to the absence of the contact interaction $i \chi_+ \partial_\mu \chi_- \bar t_L \gamma^\mu t_L  / v^2 + \textrm{h.c.}$ when $\cunoL + \ctreL = 0$, see App.~\ref{App:chiL}. 

\begin{figure}[t]
 \begin{center}
\includegraphics[width=0.6\textwidth]{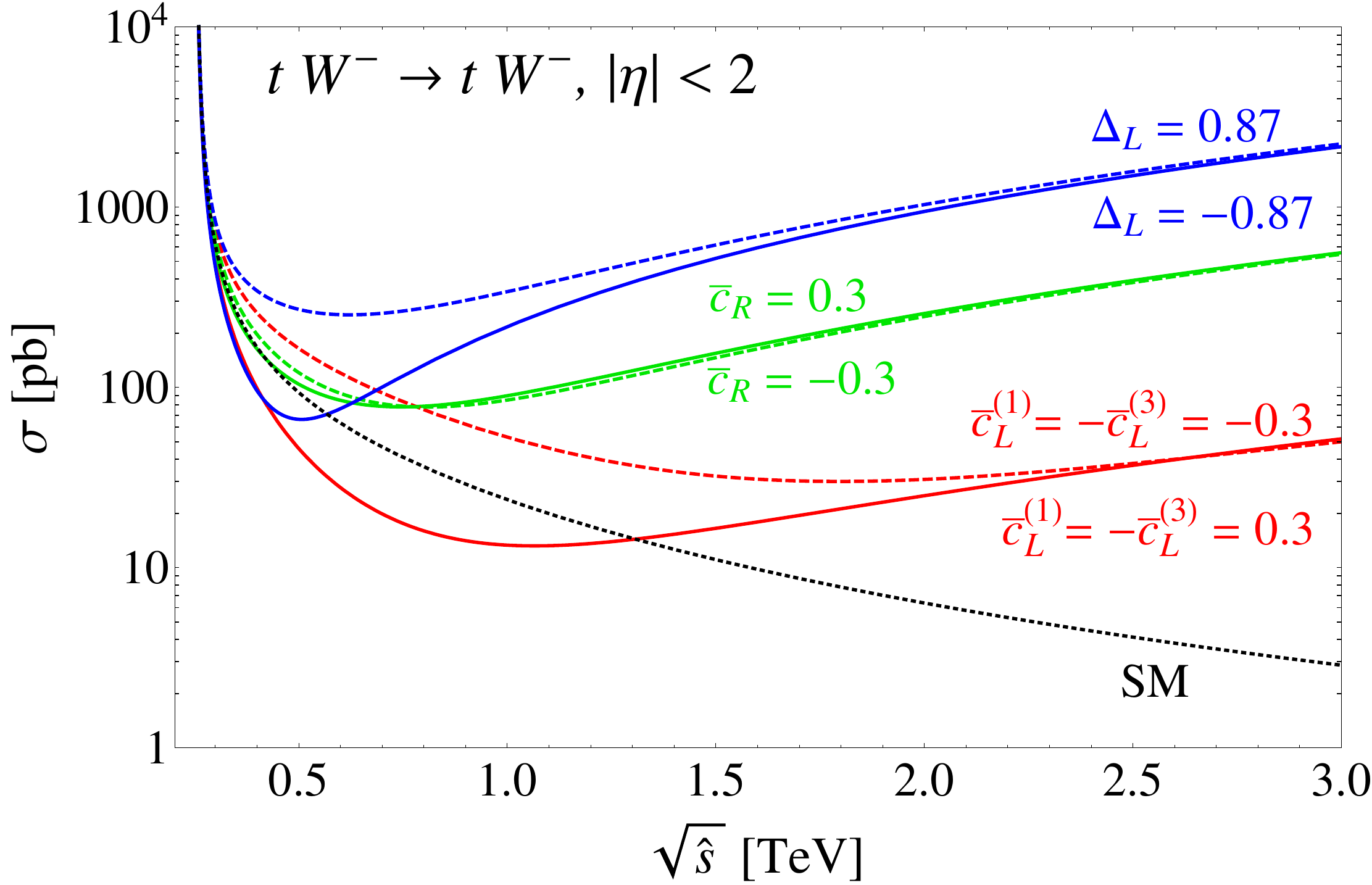}
 \end{center}
 \caption{Partonic cross section for the process $t W^{-}\to t W^{-}$ as a function of the center of mass energy $\sqrt{\hat{s}}$. The values of $\Delta_{L}$ and $\bar{c}_L^{(1)} = -\bar{c}_L^{(3)}$ are chosen to obtain the same $Zt_L t_L$ coupling for the blue and red solid curves ($\Delta_{L} < 0$) and for the blue and red dashed curves ($\Delta_{L} > 0$). For the $Zt_R t_R$ coupling there is a one-to-one correspondence between $\cR$ and $\Delta_R$, so we show only one set of curves. A pseudorapidity cut $\left|\eta\right|<2$ has been applied to remove the forward singularity, whereas the soft singularity $\hat{s}\to (m_{W}+m_{t})^{2}$ is evident from the plot. Both singularities arise due to the diagram where a photon is exchanged in the $t$-channel. At large energy, the red, blue and green curves diverge like $\hat{s}$, whereas the SM cross section (dotted black) falls off as $1/\hat{s}$.}
\label{fig:tW_tW}
\end{figure}

The cross section for $tW \to tW$ scattering is shown in Fig.~\ref{fig:tW_tW}, assuming representative values of the parameters $(\Delta_L,\Delta_R)$ and $(\cL, \cR)$.
%
%
As we already discussed, while there is a one-to-one correspondence between $\Delta_R$ and $\cR$, the coupling and HDO hypotheses genuinely differ in the left-handed interactions, because in the HDO case the $Wt_L b_L$ vertex is also modified. To facilitate the comparison, in Fig.~\ref{fig:tW_tW} we choose values of $\Delta_{L}$ and $\bar{c}_{L}$ that yield the same $Zt_L t_L$ coupling. The resulting difference is striking: for $\cL\neq 0$, the cross section is strongly suppressed compared to the case where $\Delta_L \neq 0$. This is mainly due to the cancellation of the $O(\cL)$ piece in the leading amplitude, see Eq.~\eqref{ampEFT}, which implies that the leading term of the cross section is $O(\cL^{\,4})$. This in turn translates into a weaker sensitivity to $\cL$ with respect to $\Delta_L$, because the latter appears in the leading term of the cross section at $O(\Delta_L^2)$. Additionally, from Fig.~\ref{fig:tW_tW} we learn that the cross section is enhanced for all energies, compared to the SM, if $\Delta_L > 0\,(\cL < 0)$, while for the opposite sign it is actually suppressed at low values of $\sqrt{\hat{s}}$. Once the LHC parton luminosities are taken into account, we thus expect a weaker sensitivity to the region with $\Delta_L < 0\,(\cL > 0)$. The effect is particularly striking for $\cL > 0$, in which case the cross section becomes larger than the SM one only well above $1$ TeV. These preliminary considerations, which were derived by simple inspection of the cross section of the hard scattering process $tW\to tW$, will find confirmation in the results presented below.

We now turn to the discussion of the $pp\to t\bar{t}Wj$ process at the LHC. In the following we denote our signal, which arises at $O(g_s g_w^3)$, as $(t\bar{t}Wj)_{\mathrm{EW}}$, to distinguish it from the leading mechanism for $t\bar{t}W$ production at the LHC, $pp\to t \bar{t}W$+jets at $O(g_s^{2+n}g_w)$ (with $n\geq 0$ the number of jets), which we denote as $(t\bar{t}W$+jets$)_{\mathrm{QCD}}$. Due to its high mass threshold, the latter process was not observed at the Tevatron, therefore the ATLAS and CMS experiments have designed searches aimed at extracting it from $8$ TeV LHC data, focusing on the SSL final state and vetoing events that contain a leptonic $Z$, to remove the contribution from $t\bar{t}Z$ production. The main background is constituted by processes (mostly $t\bar{t}$+jets) giving \emph{misidentified leptons} (misID$\ell$), which primarily arise from the decay of heavy flavor hadrons. The latest searches \cite{ATLASttV,CMSmva} make use of multivariate techniques and thus cannot be straightforwardly reinterpreted, but the CMS cut-and-count analysis \cite{CMSttV} contains all the information required to set a first bound on top-$Z$ interactions by exploiting the growth with energy of the $(t\bar{t}Wj)_{\mathrm{EW}}$ process. While this search was not optimized for our signal, we will use it to demonstrate the effectiveness of our approach. It is important to notice that since $(t\bar{t}Wj)_{\mathrm{EW}}$ is formally of higher order in the weak coupling compared to $(t\bar{t}W$+jets$)_{\mathrm{QCD}}$, it was neglected by CMS in the SSL analysis of Ref.~\cite{CMSttV}. Thus we perform a Monte Carlo (MC) simulation of the signal and apply the cuts chosen by CMS, obtaining the number of events expected in $8$ TeV data as function of the parameters $(\Delta_L, \Delta_R)$ or $(\cL, \cR)$. We find   
\begin{align} \label{8TeVyieldcoupl}
N_{(t\bar{t}Wj)_{\mathrm{EW}}}(\Delta_L, \Delta_R) \,=&\,\, 1.6 + 1.0\, \Delta_L + 4.1\, \Delta_L^2 + 0.3\, \Delta_R +  1.1\, \Delta_L \Delta_R + 1.0\, \Delta_R^2\,, \\\label{8TeVyieldeft}
N_{(t\bar{t}Wj)_{\mathrm{EW}}} (\cL, \cR) \,=&\,\, 1.6 - 6.2\, \cL + 8.7\, \cL^{\,2} - 7.0\, \cL^{\,3} + 11.2\, \cL^{\,4} + 0.8\, \cR \nonumber\\\,&\,\qquad\qquad\qquad\qquad\;\;\, - 2.1\, \cL \cR - 4.1\, \cL^{\,2} \cR + 10.3\, \cR^{\,2}\,.
\end{align}
Notice that the cross section is a polynomial of second order in the coupling deviations $\Delta_{L,R}$, whereas in the HDO case it is of quartic order, because two $\cL$ insertions are possible in the diagram with $b$-exchange, see Fig.~\ref{fig:FeynmanDiagrams}. Inspecting Eq.~(\ref{8TeVyieldcoupl}) (Eq.~(\ref{8TeVyieldeft})), we confirm that the pure new physics contributions, which according to the expressions of the leading amplitudes in Eq.~(\ref{ampcoup}) (Eq.~(\ref{ampEFT})) are proportional to $\Delta_{L}^2, \Delta_{R}^2$ ($\cL^{\,4}, \cR^{\,2}$), dominate over the interference and the SM terms. In addition, based on the form of the leading amplitudes we expect the following relations to hold approximately: the ratio of the coefficients of $\Delta_L^2$ ($\cL^{\,4}$) to $\Delta_R^2$ ($\cR^{\,2}$) should be equal to $\left[1- 3/(4 s _w ^2)\right]^2 (1)$. These equalities are indeed satisfied within $15\%$. 

\begin{figure}[t]
 \begin{center}
\includegraphics[width=0.36\textwidth]{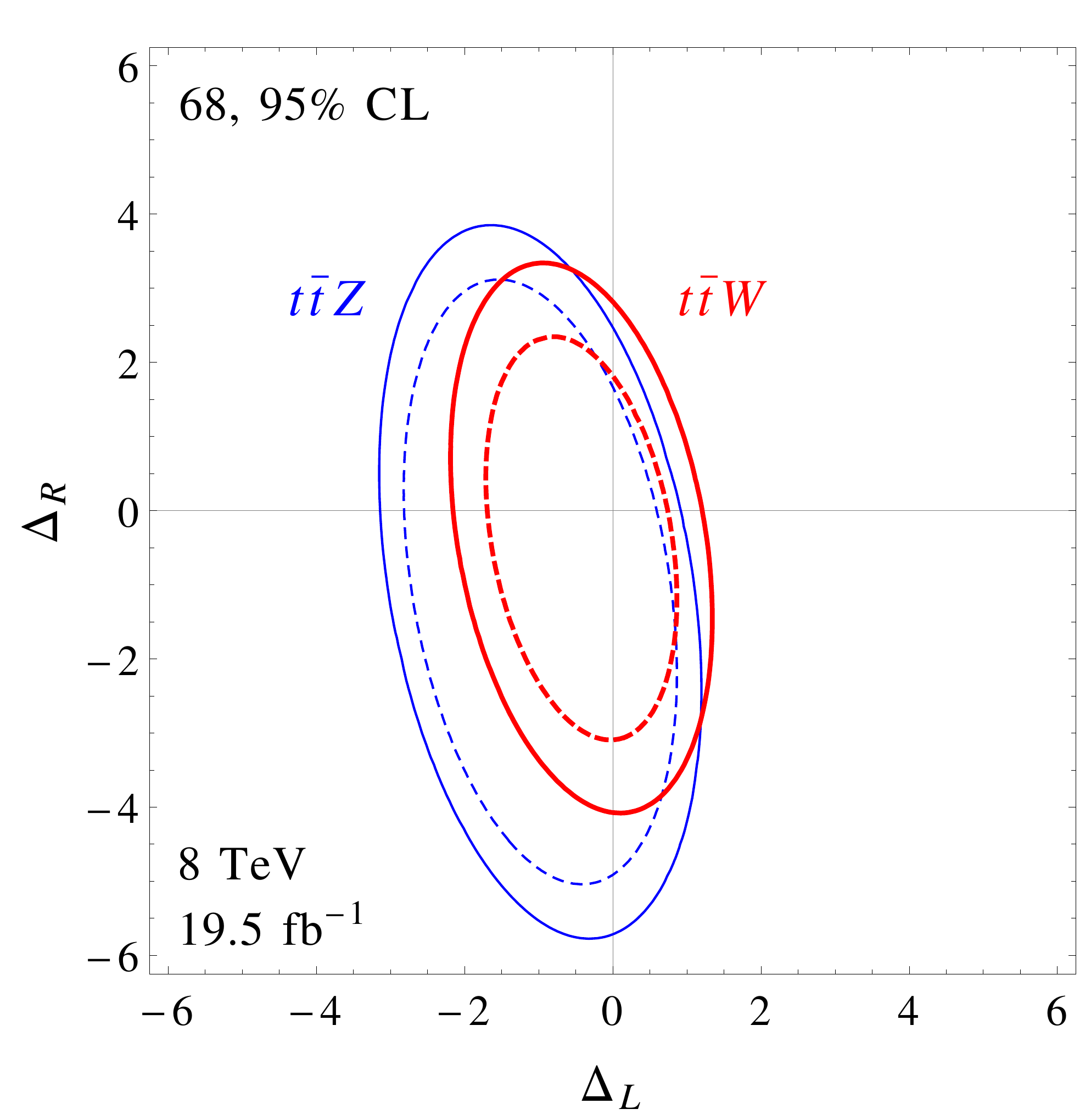}
\includegraphics[width=0.36\textwidth]{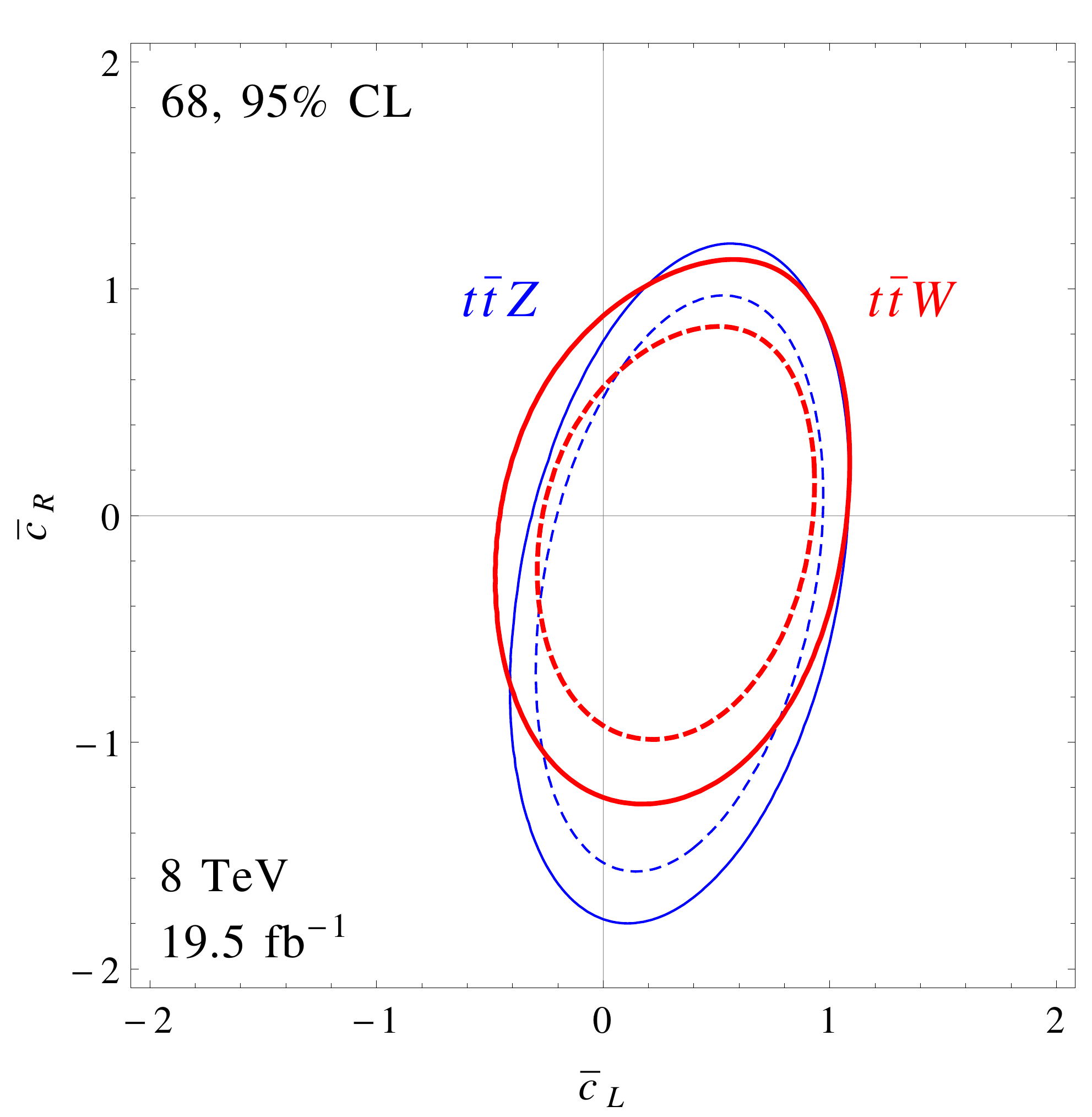}
 \end{center}
 \caption{In red, the constraints on top-$Z$ coupling deviations (left panel) and HDO coefficients (right panel) derived from the $t\bar{t}W$ analysis at $8$ TeV. For comparison, in blue we show the constraint obtained from the $8$ TeV $t\bar{t}Z$ analysis.}
\label{fig:8TeVbounds}
\end{figure}

For comparison, CMS quotes an expected yield of $14.5$ events for $(t\bar{t}W+\mathrm{jets})_{\mathrm{QCD}}$. Thus from Eqs.~(\ref{8TeVyieldcoupl}, \ref{8TeVyieldeft}) we see that while in the SM $(t\bar{t}Wj)_{\mathrm{EW}}$ only provides a $\sim 10\%$ correction to the $(t\bar{t}W+\mathrm{jets})_{\mathrm{QCD}}$ yield, it grows rapidly moving away from the SM point. This, together with the fact that CMS did not observe any excess over the SM expectation, allows us to set a bound on $\Delta_{L,R}$ or $\bar{c}_{L,R}$. The results are shown in Fig.~\ref{fig:8TeVbounds}, where for comparison we also display the bounds obtained from the $t\bar{t}Z$ CMS analysis in the trilepton final state \cite{CMSttV}, which according to common wisdom provides the best constraint on top-$Z$ couplings at the LHC. Strikingly, we find that the best current constraints are instead provided by the $t\bar{t}W$ channel, so far thought to be insensitive to top-$Z$ interactions.  
%
This result becomes even more remarkable when we consider that the CMS analysis was optimized to increase the sensitivity not to our signal, but to the main irreducible background $(t\bar{t}W$+jets$)_{\mathrm{QCD}}$. Inspecting the HDO bound in the right panel of Fig.~\ref{fig:8TeVbounds}, we note that the coefficients of the dim-$6$ operators are allowed to be of $O(1)$. Thus the interpretation of the result in terms of HDO is not truly justified, and should be intended as purely illustrative of the current sensitivity. Assuming only a modification of the $Zt_R t_R$ coupling, we find for $8 \TeV$, $19.5\;\mathrm{fb}^{-1}$ at $95\%$ CL\footnote{Given the very large $Ztt$ coupling deviations allowed by $8$ TeV data, one may wonder about effects in the $t\bar{t}$ forward-backward asymmetry measured at the Tevatron. The tree-level contribution due to $q\bar{q}\to Z,\gamma\to t\bar{t}$ is $\sim 0.2\%$ in the SM \cite{HollikPagani}, and we estimate that, within the allowed region shown in the left panel of Fig.~\ref{fig:8TeVbounds}, it is enhanced by a factor $\lesssim 5$, thus remaining strongly subdominant to the QCD contribution, which amounts to approximately $8\%$ \cite{HollikPagani}. Interestingly, at the LHC the $t\bar{t}$ charge asymmetry in the $t\bar{t}W$ process is significantly larger than in inclusive $t\bar{t}$ production \cite{ttWchargeasym}.}
\begin{equation}\label{ZtRtRonedim8}
\qquad-3.6 < \Delta_R < 2.4\qquad \mathrm{or}\qquad -1.13 < \cR < 0.74\,.
\end{equation}   

\begin{figure}[t!]
 \begin{center}
\includegraphics[width=0.325\textwidth]{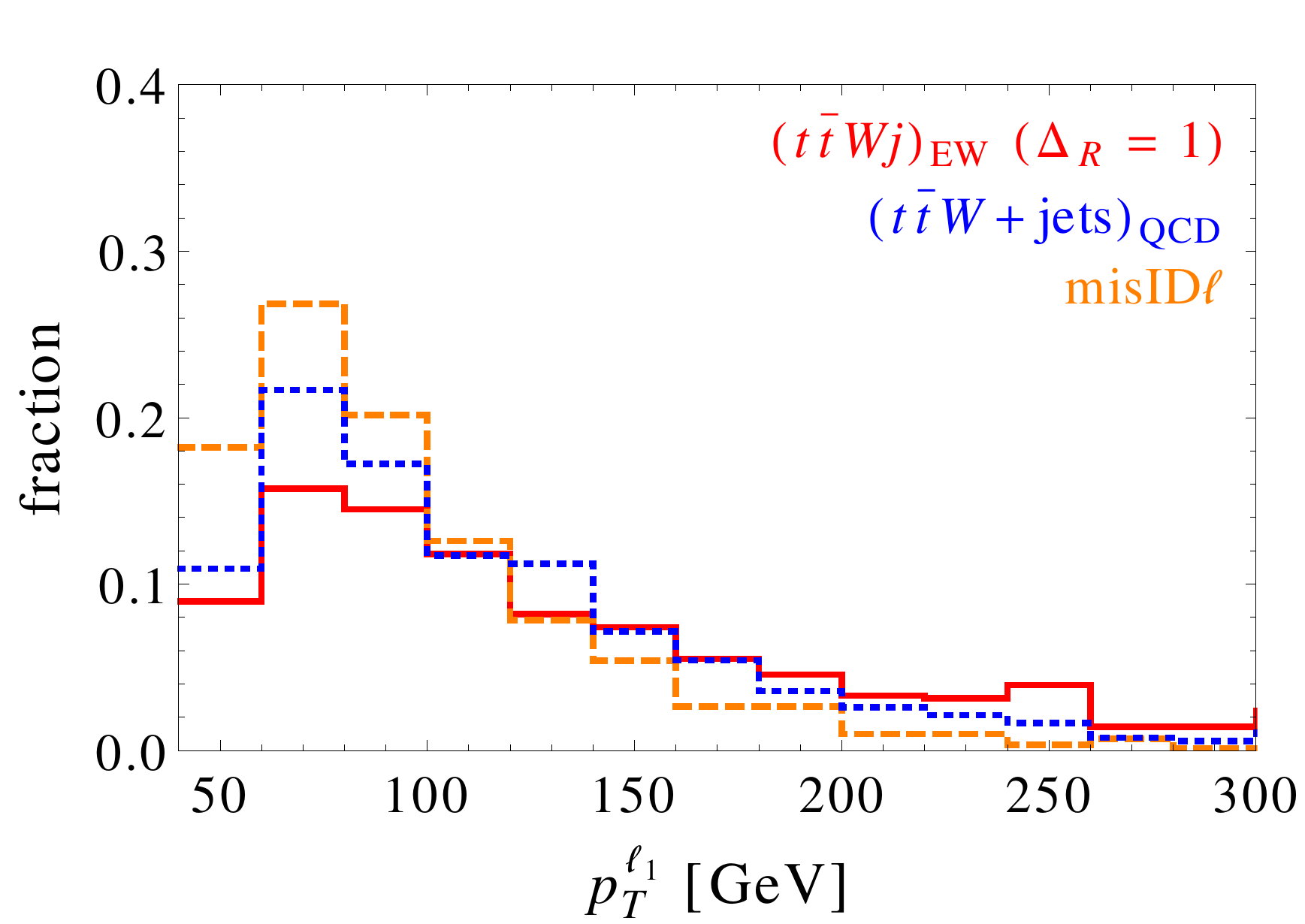}
\includegraphics[width=0.325\textwidth]{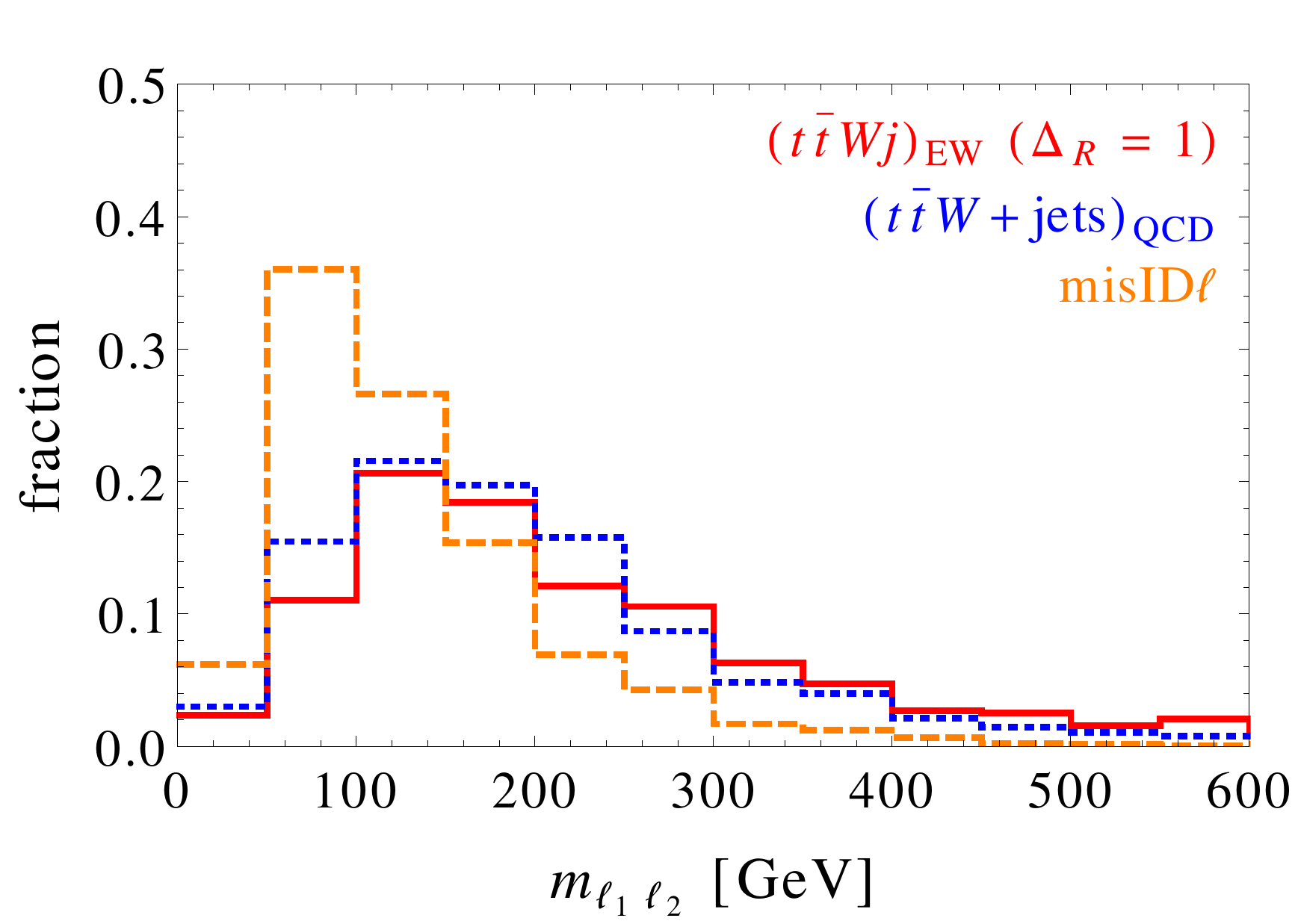}
\includegraphics[width=0.325\textwidth]{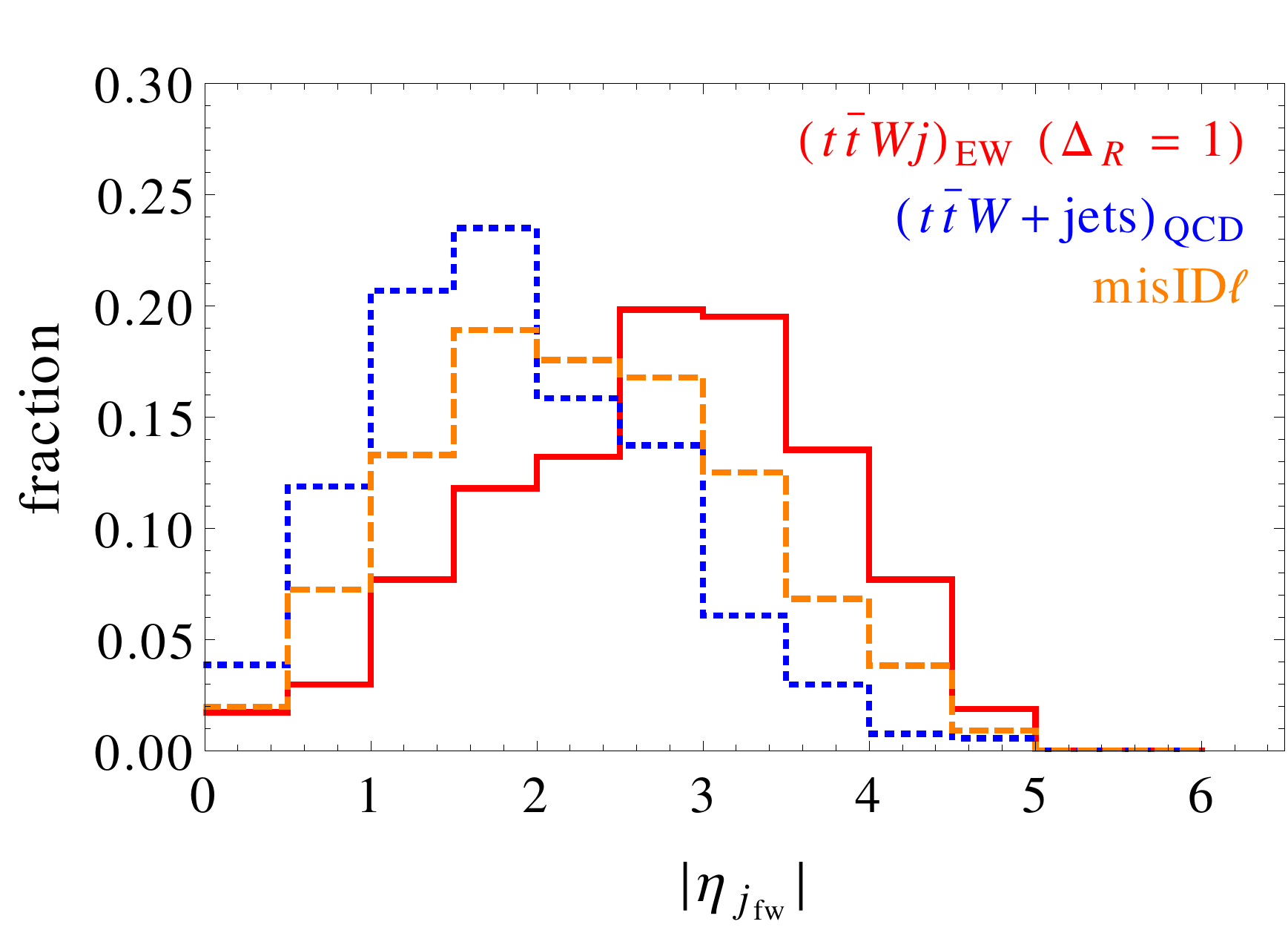}
 \end{center}
 \caption{Normalized distributions for the signal $(t\bar{t}Wj)_{\mathrm{EW}}$ and the two main backgrounds $(t\bar{t}W+\mathrm{jets})_{\mathrm{QCD}}$ and misID$\ell$ at $13$ TeV, after the $4j$ pre-selection.}
\label{fig:13TeVdistribsec3}
\end{figure}

Having proven the effectiveness of our method, we move on to designing a search at $13$ TeV that specifically targets the process $(t\bar{t}Wj)_{\mathrm{EW}}$. The latter has two distinctive features that can be exploited to separate it from the background: a $tW$ pair with large invariant mass (where $t$ can be either top or antitop, and $W$ either of $W^\pm$), due to the growth with energy of the hard scattering process, and a highly energetic forward jet arising from the radiation of a $W$ off an initial-state quark. We devise cuts that single out events with these properties and thus increase the significance of the signal over the background, which is mainly composed by $(t\bar{t}W$+jets$)_{\mathrm{QCD}}$ and misID$\ell$. We validate our background simulations against the CMS $8$ TeV results, and perform the cut optimization using the point $(\Delta_L, \Delta_R) = (0,1)$ as signal benchmark. This choice is motivated by the fact that the $Zt_R t_R$ coupling is currently very weakly constrained even under the assumption of heavy new physics, in contrast with the $Zt_L t_L$ coupling, which within the HDO framework is already bounded by the measurements of $Z b_L b_L$ and of $W t_L b_L$. Our basic selection requires two SSL and $\geq4$ jets, among which $\geq 1$ must be $b$-tagged. We identify a set of useful kinematic variables to enhance the significance of the signal, which are discussed in detail in Sec.~\ref{4janalysis}. For illustration, in Fig.~\ref{fig:13TeVdistribsec3} we show the normalized distributions of signal and backgrounds for a subset of these variables: the transverse momentum of the leading lepton, $p_T^{\ell_1}$, the invariant mass of the two leading leptons, $m_{\ell_1 \ell_2}$, and the pseudorapidity of the forward jet, $|\eta_{j_{\mathrm{fw}}}|$.
%
%
It is apparent that the leptonic variables are effective in suppressing the misID$\ell$ background, whereas a lower cut on the pseudorapidity of the forward jet helps to suppress $(t\bar{t}W$+jets$)_{\mathrm{QCD}}$. The event yields after all cuts, assuming an integrated luminosity of $300\;\mathrm{fb}^{-1}$, are given by
\begin{align} \label{13TeVyieldcoupl4j}
N_{(t\bar{t}Wj)_{\mathrm{EW}}}(\Delta_L, \Delta_R) \,=&\,\, 16.9 + 12.7\, \Delta_L + 172.4\, \Delta_L^2 + 0.5\, \Delta_R + 37.2\, \Delta_L \Delta_R + 40.8\, \Delta_R^2\,, \\\label{13TeVyieldhdo4j}
N_{(t\bar{t}Wj)_{\mathrm{EW}}} (\cL, \cR) \,=&\,\, 16.7 - 73.2 \,\cL + 145.0\, \cL^{\,2} - 164.2\, \cL^{\,3} + 
 408.3\, \cL^{\,4} + 6.3\, \cR \nonumber\\ \,&\,\qquad\qquad\qquad\qquad\qquad\;\;\, 
- 4.1\, \cL \cR - 121.8\, \cL^{\,2} \cR + 
 412.3\, \cR^{\,2}\,.
\end{align} 

\begin{figure}[t!]
 \begin{center}
\includegraphics[width=0.37\textwidth]{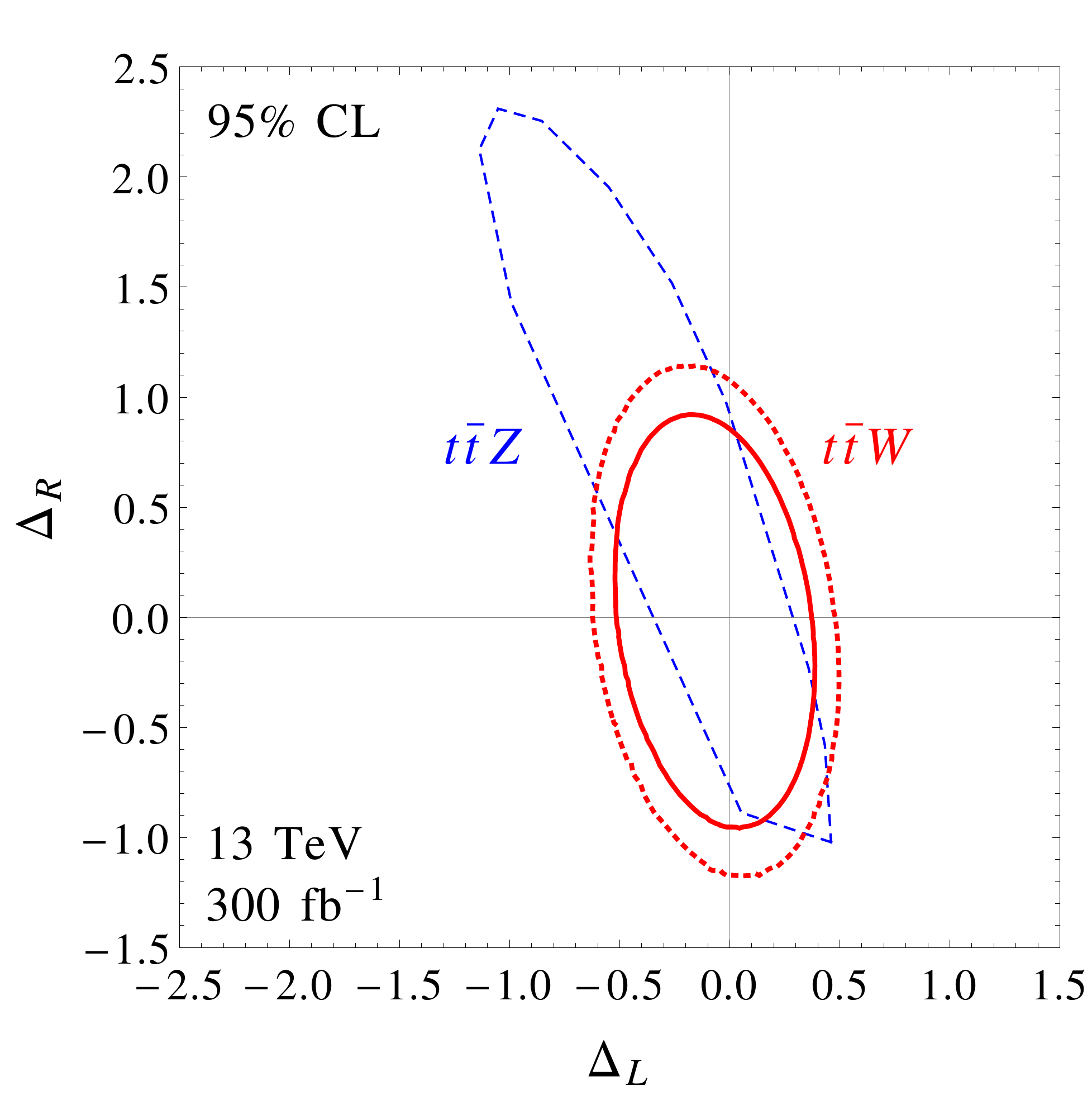}
\includegraphics[width=0.37\textwidth]{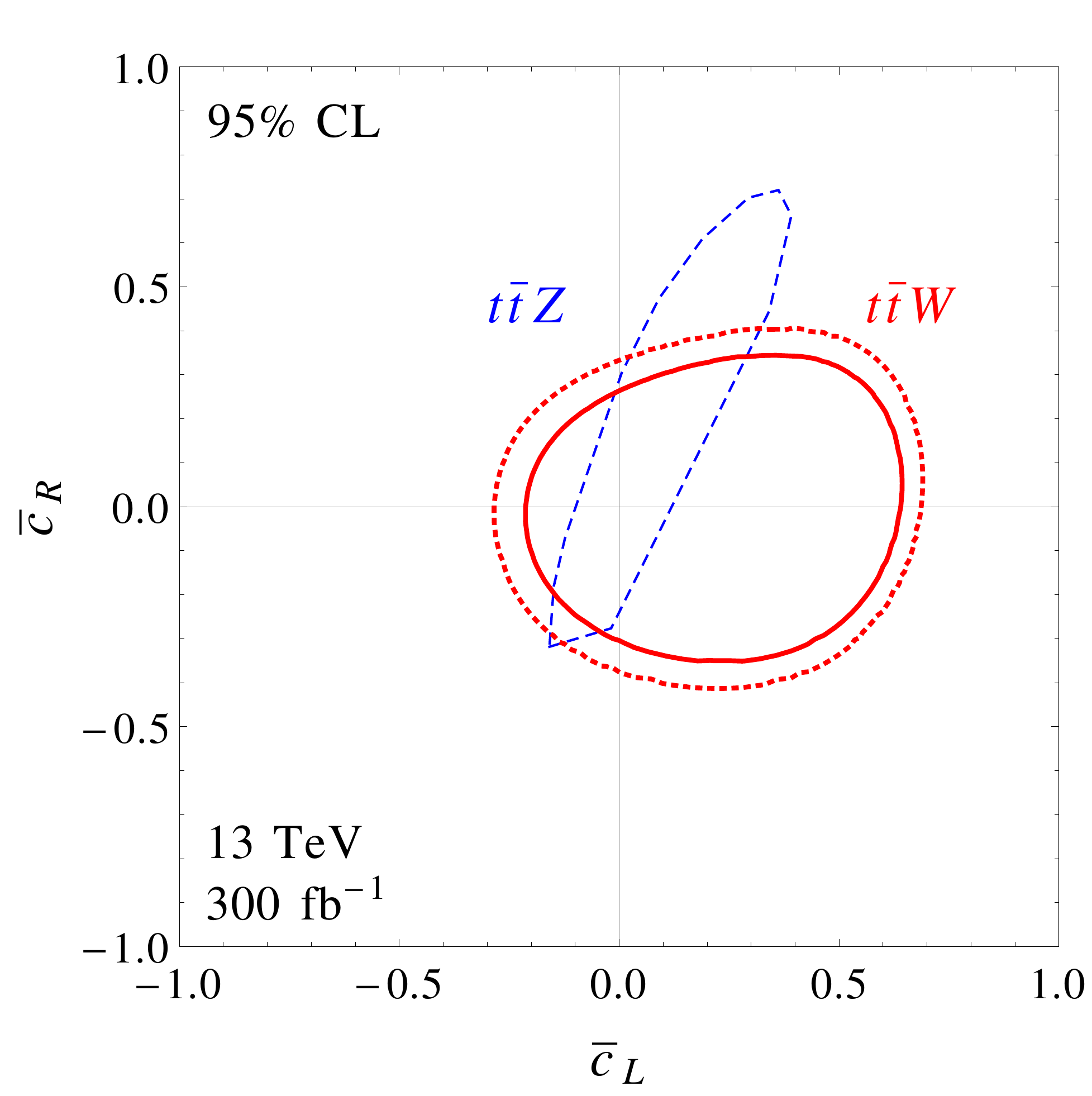}\\
 \end{center}
 \caption{In red, the constraints on top-$Z$ coupling deviations (left panel) and HDO coefficients (right panel) derived from our $4j$ $t\bar{t}W$ analysis at $13$ TeV. The solid contour assumes no systematic uncertainty on the background, whereas the dotted one includes a $50\%$ systematic on the misID$\ell$ component. For comparison, in dashed blue we show the constraint obtained from $t\bar{t}Z$, as derived in Ref.~\cite{RS} by means of a NLO-QCD signal-only analysis.}
\label{fig:13TeVbound_4j}
\end{figure}

The expected background yield is of $51$ events for $(t\bar{t}W+\mathrm{jets})_{\mathrm{QCD}}$, and of $34$ events for misID$\ell$. By performing a simple likelihood analysis, we obtain the constraints on $\Delta_{L,R}$ and $\bar{c}_{L,R}$ shown as red contours in Fig.~\ref{fig:13TeVbound_4j}. 
%
%
The solid contours assume no systematic uncertainty on the background, whereas the dotted contours include the dominant $50\%$ systematic uncertainty on the misID$\ell$ component. For comparison, in the same figure we show the projected $13$ TeV bounds from the $t\bar{t}Z$ process, as derived in Ref.~\cite{RS}. This comparison is meant to be illustrative, because the projection of Ref.~\cite{RS} is based on a NLO-QCD analysis of the signal, without the inclusion of detector effects nor backgrounds. The two main effects that were gleaned by inspecting the partonic cross section in Fig.~\ref{fig:tW_tW} are now manifest in Fig.~\ref{fig:13TeVbound_4j}. First, the sensitivity to $\cL$ is weaker than to $\Delta_L$, because the former appears in the leading term of the cross section at $O(\cL^4)$ while the latter at $O(\Delta_L^2)$, see Eqs.~(\ref{ampcoup}) and (\ref{ampEFT}). Second, the $\cL > 0$ direction is less strongly constrained than $\cL < 0$, because in the former case the partonic cross section for $tW\to tW$ scattering is smaller than the SM one for $\sqrt{\hat{s}}\lesssim \mathrm{TeV}$, where the bulk of the LHC parton luminosity is concentrated. Comparing with the $t\bar{t}Z$ process, we find that our $t\bar{t}W$ analysis gives a significantly stronger bound on the coupling deviations $\Delta_{L,R}$, and comparable sensitivity to the HDO coefficients $\bar{c}_{L,R}$. We also note that in the HDO case the shape of the $t\bar{t}W$ contours is rather different from that of the $t\bar{t}Z$ ones, leading to an interesting complementarity of the two measurements. Assuming only a modification of the $Zt_R t_R$ coupling, our analysis gives for $13 \TeV$, $300\;\mathrm{fb}^{-1}$ at $95\%$ CL    
\begin{align}
\qquad-0.83 < \Delta_R < 0.74\qquad\, &\mathrm{or}\qquad -0.26 < \cR < 0.23\ ,
\label{ZtRtRonedim13}
\end{align}   
with no systematics on the background, while if a $50\%$ systematic uncertainty on the misID$\ell$ component is included, we find $-1.04 < \Delta_R < 0.95$ or $-0.32 < \cR < 0.30$. 
Based on these results, we urge the ATLAS and CMS collaborations to search for the $(t\bar{t}Wj)_{\mathrm{EW}}$ signal in the upcoming $13$ TeV data. The necessary technical details of our analysis are presented in the next section, which the reader interested only in the discussion of our results can omit, to move directly to Sec.~\ref{sec:OtherProcesses}.  

\section{$t\bar{t}W$ analysis}\label{sec:analysis}
In this section we present the technicals details of our analysis. Frequent reference will be made to the $8$ TeV CMS cut-and-count search for $t\bar{t}W$ \cite{CMSttV}, based on the requirements of two SSL and a leptonic $Z$ veto. After reinterpreting this search to obtain the $8$ TeV bounds on top-$Z$ interactions, we employ its results to validate our background simulations. We then propose a dedicated $13$ TeV analysis that targets the signal process $(t\bar{t}Wj)_{\mathrm{EW}}$.    
\subsection{8 TeV bounds}
The SSL analysis of Ref.~\cite{CMSttV} was aimed at measuring the $(t\bar{t}W$+jets$)_{\mathrm{QCD}}$ process, while our signal process $(t\bar{t}Wj)_{\mathrm{EW}}$ was neglected. On the other hand, the $(t\bar{t}Wj)_{\mathrm{EW}}$ amplitude interferes with the one-jet component of $(t\bar{t}W$+jets$)_{\mathrm{QCD}}$, which we will label $(t\bar{t}Wj)_{\mathrm{QCD}}$, thus {\it a priori} our signal cannot be generated separately from the $(t\bar{t}W$+jets$)_{\mathrm{QCD}}$ process. A further subtlety arises because the $t\bar{t}W$ final state can also be produced purely from weak interactions, at $O(g_w^3)$. To quantify these effects, we compute inclusive parton-level cross sections for the SM and one representative signal point, which is chosen to be $\Delta_R =3.2$ at $8$ TeV and $\Delta_R = 1$ at $13$ TeV, roughly corresponding to the sensitivity of our analysis (see Figs.~\ref{fig:8TeVbounds} and \ref{fig:13TeVbound_4j}, respectively). The cross sections are computed with MadGraph5 \cite{MG5}, employing a FeynRules \cite{Feynrules} model that allows us to add to the SM either the corrections $\Delta_{L,R}$ to the top-$Z$ couplings, or the dim-$6$ operators proportional to $\bar{c}_L^{\,(1)},\bar{c}_L^{\,(3)},\bar{c}_R$. The model was validated against analytical computations of several $2\to 2$ amplitudes, and employed for all the MC simulations used in this paper. For the SM parameters we take the values
\begin{align}
m_Z = 91.19\;\mathrm{GeV}\, ,\qquad &\alpha(m_Z)=1/127.9\,,\qquad G_F = 1.166\times 10^{-5}\;\mathrm{GeV}^{-2}\,, \nonumber\\
\alpha_s (m_Z) &= 0.1184\,,\qquad m_t = 173\;\mathrm{GeV}\,.
\end{align}
Inspection of the inclusive cross sections in Table~\ref{tab:InterfParton} shows that the pure electroweak contribution to $t\bar{t}W$ is very small, thus we will neglect it in our study. On the other hand, the effect of the interference between the $(t\bar{t}Wj)_{\mathrm{QCD}}$ and $(t\bar{t}Wj)_{\mathrm{EW}}$ amplitudes on the deviation from the SM cross section in presence of anomalous top-$Z$ couplings is at most $20\%$. Given the exploratory nature of our study, for simplicity we choose to perform our analysis neglecting the interference, and take into account its effect by including a conservative $20\%$ systematic uncertainty on the $(t\bar{t}Wj)_{\mathrm{EW}}$ signal. 
\begin{table}[t]
   \begin{center}
   \begin{tabular}{cc|c c|c c c c} 
    & & $(t\bar{t}W)_{\mathrm{QCD}}$ & $(t\bar{t}W)_{\mathrm{EW}}$ & $(t\bar{t}Wj)_{\mathrm{QCD}}$ & $(t\bar{t}Wj)_{\mathrm{EW}}$ & $(t\bar{t}Wj)_{\mathrm{full}}$ & $1-\frac{\delta_{\mathrm{full}}}{\delta_{\cancel{\mathrm{int}}}}$ \\[0.2cm]
   \hline
   &&&& \\[-0.4cm]
   \multirow{2}{*}{$8$ TeV} & SM & $130.6$ & $0.99$ & $94.0$ & $12.6$ & $104.1$ & \multirow{2}{*}{$0.19(4)$} \\[0.02cm]
   & $\Delta_R = 3.2$ & $130.6$ & $1.73$ & $94.0$ & $64.9$ & $146.5$ & \\[0.02cm]
   \hline
   &&&& \\[-0.4cm]
   \multirow{2}{*}{$13$ TeV} & SM & $347.9$ & $2.85$ & $341.3$ & $56.0$ & $386.1$ & \multirow{2}{*}{$0.02(15)$} \\[0.02cm]
   & $\Delta_R = 1$ & $347.9$ & $2.71$ & $341.3$ & $94.6$ & $423.9$ & \\[0.02cm]
     
   \end{tabular}   
   \end{center}
   \caption{Parton-level cross sections in femtobarns. By $(t\bar{t}Wj)_{\mathrm{full}}$ we denote the full amplitude including the interference. For the $t\bar{t}Wj$ process we imposed the cuts $p_T^j > 20\;\mathrm{GeV}$ and $\left|\eta\right| < 5$. The quantity $\delta_{\mathrm{full},\,\cancel{\mathrm{int}}} \equiv \sigma^{\Delta_R\neq 0}_{(t\bar{t}Wj)_{\mathrm{full},\,\mathrm{EW}}}-\sigma^{\mathrm{SM}}_{(t\bar{t}Wj)_{\mathrm{full},\,\mathrm{EW}}}$ is the deviation from the SM, computed either including (`full') or neglecting (`\cancel{int}') the interference. In the last column, the uncertainty in parentheses refers to the last digit.}
   \label{tab:InterfParton}
\end{table}

Because we neglect the interference, to compute the constraints on top-$Z$ interactions we need to apply the CMS cuts to the $(t\bar{t}Wj)_{\mathrm{EW}}$ process, and extract the dependence of the signal event yield on the parameters $\Delta_{L,R}$ and $\bar{c}_{L,R}$. The signal yield will then be summed to those of the processes already simulated in Ref.~\cite{CMSttV}, including $(t\bar{t}W+\mathrm{jets})_{\mathrm{QCD}}$. Signal events are generated with MadGraph5, employing our FeynRules model. Showering and hadronization effects are accounted for with Pythia 6.4 \cite{Pythia6}, and the detector simulation is performed using PGS4 \cite{PGS}. To match Ref.~\cite{CMSttV}, the following changes are made to the default CMS settings in PGS: the $b$-tagging is modified to reproduce the performance of the medium working point of the CSV algorithm, and the jet reconstruction algorithm is set to anti-$k_T$ with distance parameter of $0.5$. In addition, the calorimeter coverage for jets is extended up to $|\eta| = 5$. We make use of NN23LO1 parton distribution functions \cite{NNPDF}, and factorization and renormalization scales are set to the default MadGraph5 event-by-event value. Unless otherwise noted, the above settings are used for all the event samples used in this paper. The event selection requirements follow closely those listed in Sec.~4 of Ref.~\cite{CMSttV}, and are as follows:
\begin{enumerate}
\item Two SSL, each with $|\eta| < 2.4$ and $p_T > 40\;\mathrm{GeV}$;
\item At least three jets with $|\eta| < 2.4$ and $p_T > 30\;\mathrm{GeV}$, among which at least one must be $b$-tagged;
\item An event is rejected if it contains, in addition to the SSL pair, $2$ or more leptons with $|\eta| < 2.4$ and $p_T > 10\;\mathrm{GeV}$, or if it contains one such lepton forming, with one of the two SSL, a same-flavour opposite-sign pair whose invariant mass is within $15$ GeV of $m_Z$;
\item $H_T > 155\;\mathrm{GeV}$, where $H_T$ is the scalar sum of the transverse momenta of all jets as defined in point 2;
\item The CMS lepton isolation is approximated by requiring that $\Delta R (\ell, j) > 0.3$ for each of the SSL and for all jets as defined in point 2. 
\end{enumerate}  
The events are divided in $6$ categories depending on the flavor/charge combination of the SSL. The expected event yields for all the processes considered in Ref.~\cite{CMSttV}, after summing over all SSL categories, are shown in Table~\ref{tab:8TeVyields}.
\begin{table}[t]
   \begin{center}
   \begin{tabular}{c c c c c c|c|c} 
   $(t\bar{t}W$+jets$)_{\mathrm{QCD}}$ & misID$\ell$ & irreducible & $t\bar{t}Z$ & misID$Q$ & $WZ$ & total & observed \\[0.02cm]
   \hline
   &&&&&&& \\[-0.4cm]
   $14.5$ & $12.1$ & $5.8$ & $3.9$ & $2.2$ & $1.3$ & $39.8$ & $36$ \\[0.02cm]
   \end{tabular}   
   \end{center}
   \caption{Expected and observed background yields for the $8$ TeV SSL analysis, after summing over all SSL categories. The numbers are taken from Ref.~\cite{CMSttV}.}
   \label{tab:8TeVyields}
\end{table}
The largest SM contribution is given by $(t\bar{t}W$+jets$)_{\mathrm{QCD}}$, which was considered as signal in Ref.~\cite{CMSttV}, but will be a background in our analysis. The second contribution is given by the misID$\ell$ background, composed of processes with one prompt and one non-prompt lepton. The latter arises from the decay of a heavy flavor hadron, and is misidentified as prompt. The misID$\ell$ background is dominated by $t\bar{t}$ events. Subleading contributions are given by the `irreducible' processes, which include $t\bar{t}h$ and same-sign $WW$ production in association with jets, and by $t\bar{t}Z$. A minor background is given by processes where the misidentification of the charge of one electron leads to the SSL final state. This contribution, dominated by $t\bar{t}$ and Drell-Yan (DY)$+\mathrm{jets}$ events, is labeled {\it misidentified charge} (misID$Q$) background. Finally, $WZ$+jets production is also a minor background. 

To efficiently compute the signal yield after cuts as function of the parameters $\Delta_{L,R}$ and $\bar{c}_{L,R}$, we exploit the fact that formally the $(t\bar{t}Wj)_{\mathrm{EW}}$ cross section is a polynomial of second order in $\Delta_{L,R}$, and of quartic order in $\bar{c}_{L,R}$. 
Thus it is sufficient to generate a small number of signal samples and perform a fit, which yields semi-numerical formulas parameterizing the signal predictions. For brevity, only the sum over all flavor/charge combinations of the SSL was reported in Eqs.~(\ref{8TeVyieldcoupl},~\ref{8TeVyieldeft}).
The statistical uncertainty on the signal yields computed using those equations is approximately $10\%$. The fact that CMS observed a number of events compatible with the SM prediction (see Table~\ref{tab:8TeVyields}) allows us to set a bound on top-$Z$ interactions. Denoting by $\vec{p}$ either $\{\Delta_L, \Delta_R\}$ or $\{\cL, \cR\}$, we thus consider the following likelihood
\begin{align}
L(\vec{p}\,; r,t) \,&=\, \prod_{i=1}^6 \frac{(N^i_{S+B})^{N^i_{\mathrm{obs}}}e^{-N^{i}_{S+B}}}{N^i_{\mathrm{obs}}!}\,P_{\sigma_r}(r,1)P_{\sigma_t}(t,1) \, , \nonumber\\
P_{\sigma}(x, x_0) \,&=\, \frac{1}{\frac{1}{2}\left[1+\mathrm{erf}\left(\frac{1}{\sqrt{2}\,\sigma}\right)\right]}\frac{1}{\sqrt{2\pi}\sigma}e^{-\frac{(x-x_0)^2}{2\sigma^2}}\, , \nonumber \\
N^i_{S+B} \,&=\, r N^i_{B,\,\mathrm{misID\ell}} + N^i_{B,\,\mathrm{other}} + N^i_{(t\bar{t}Wj)_{\mathrm{EW}}}(\vec{0}) \nonumber\\\,&\qquad\qquad\qquad+\, t(N^i_{(t\bar{t}Wj)_{\mathrm{EW}}}(\vec{p}\,)-N^i_{(t\bar{t}Wj)_{\mathrm{EW}}}(\vec{0}))\, , \label{likelihood}
\end{align}
where the dominant systematic uncertainty of $50\%$ on the misID$\ell$ background was included\footnote{We have verified that by assuming $50\%$ on the misID$\ell$ background as the only systematic uncertainty, we reproduce to good accuracy the measurement of the $t\bar{t}W$ cross section quoted in Ref.~\cite{CMSttV}: we find $178^{+106}_{-101}$ fb, to be compared with $170^{+114}_{-106}$ fb.} by setting $\sigma_r =0.5$, and we also took into account the already mentioned $20\%$ systematic uncertainty on the signal by setting $\sigma_t = 0.2$. The index $i$ runs over the $6$ SSL categories. Maximizing the marginalized log-likelihood, defined as $\ell_{m}(\vec{p}\,)=\log \left(\int_0^{+\infty} L(\vec{p}\,;r,t)dr\,dt\right)$, and taking standard confidence intervals we obtain the exclusion contours shown in Fig.~\ref{fig:8TeVbounds}. To put our constraints in perspective, we compare them with those derived from the CMS $8$ TeV $t\bar{t}Z$ analysis in the trilepton final state, also performed in Ref.~\cite{CMSttV} (see App.~\ref{App:ttZ} for details). Setting $\Delta_L =0$ (or equivalently, $\cL = 0$) in the likelihood, we obtain the one-dimensional bounds reported in Eq.~\eqref{ZtRtRonedim8}. Notice that, as shown in Table~\ref{tab:8TeVyields}, despite the leptonic $Z$ veto the $t\bar{t}Z$ process gives a small contribution to the SSL signal region. For the sake of consistency, to generate the $t\bar{t}W$ contours in Fig.~\ref{fig:8TeVbounds} we have taken into account the dependence of the $t\bar{t}Z$ event yield on the parameters $\vec{p}$, rescaling the value quoted by CMS for the SM ($3.9$ events) by the ratio $\sigma_{t\bar{t}Z}(\vec{p}\,)/\sigma_{t\bar{t}Z}^\mathrm{SM}$, with $\sigma_{t\bar{t}Z}$ the inclusive cross section for $pp\to t\bar{t}Z$ at $8$ TeV. This is based on the assumption that the selection efficiency is independent of $\vec{p}$, which is expected to be a reasonable approximation, since the leading $pp\to t\bar{t}Z$ amplitude does not grow with energy for non-SM top-$Z$ couplings. On the other hand, the subleading contribution $pp\to t\bar{t}Zj$, which is the analogue of our signal with $W \to Z$, does grow with energy, and one may wonder if it is justified to discard it. However, as discussed in Sec.~\ref{tZ_tZ}, the $tZ\to tZ$ amplitude only grows with energy as $\sqrt{\hat{s}}$, as opposed to $\hat{s}$ for $tW\to tW$. We can thus safely neglect this piece. The effect of the $t\bar{t}Z$ contamination on the $t\bar{t}W$ bounds in Fig.~\ref{fig:8TeVbounds} is small. 

\subsection{Background simulation}
\begin{figure}[t!]
 \begin{center}
\includegraphics[width=0.30\textwidth]{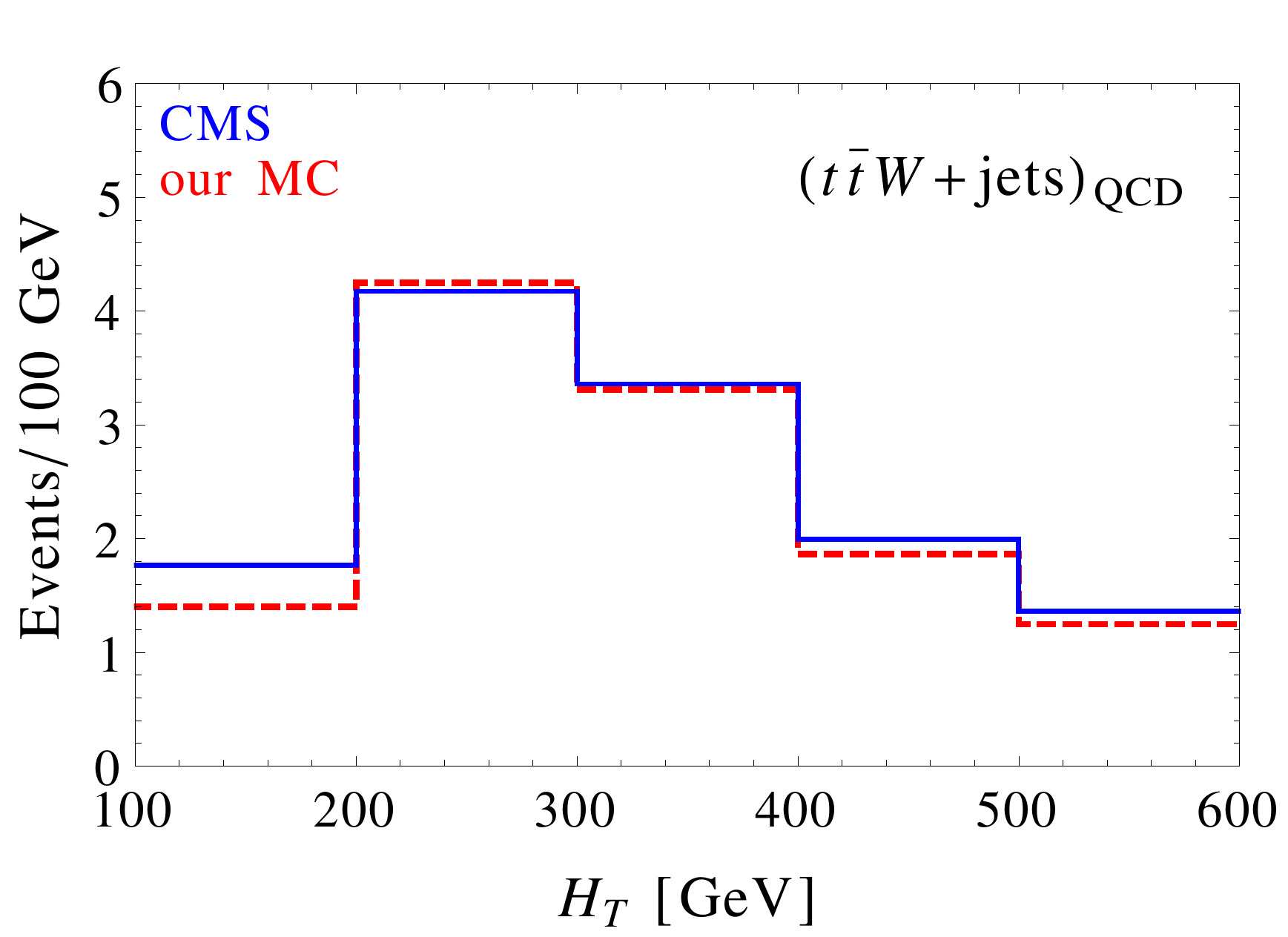}
\includegraphics[width=0.305\textwidth]{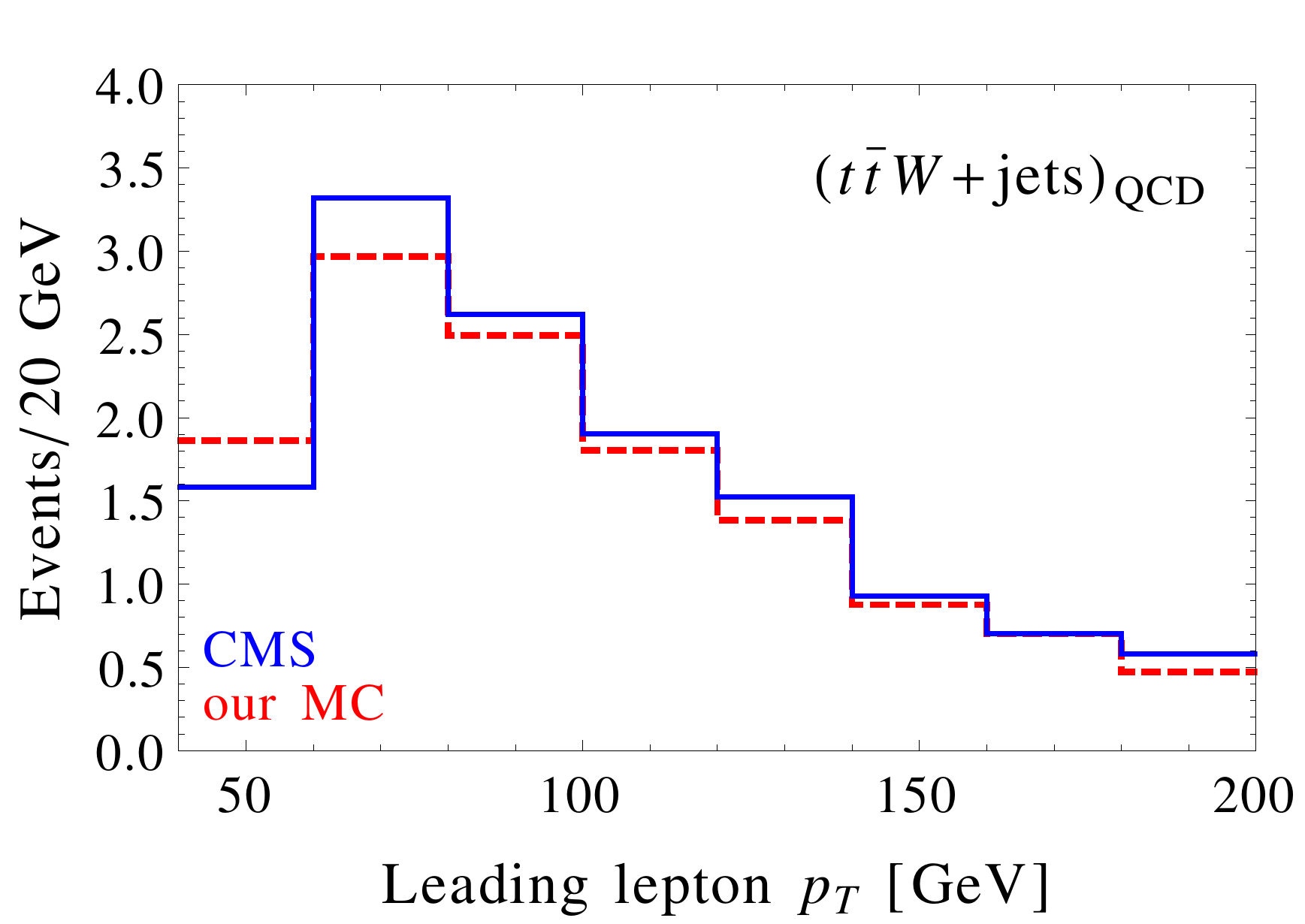}\\
\includegraphics[width=0.30\textwidth]{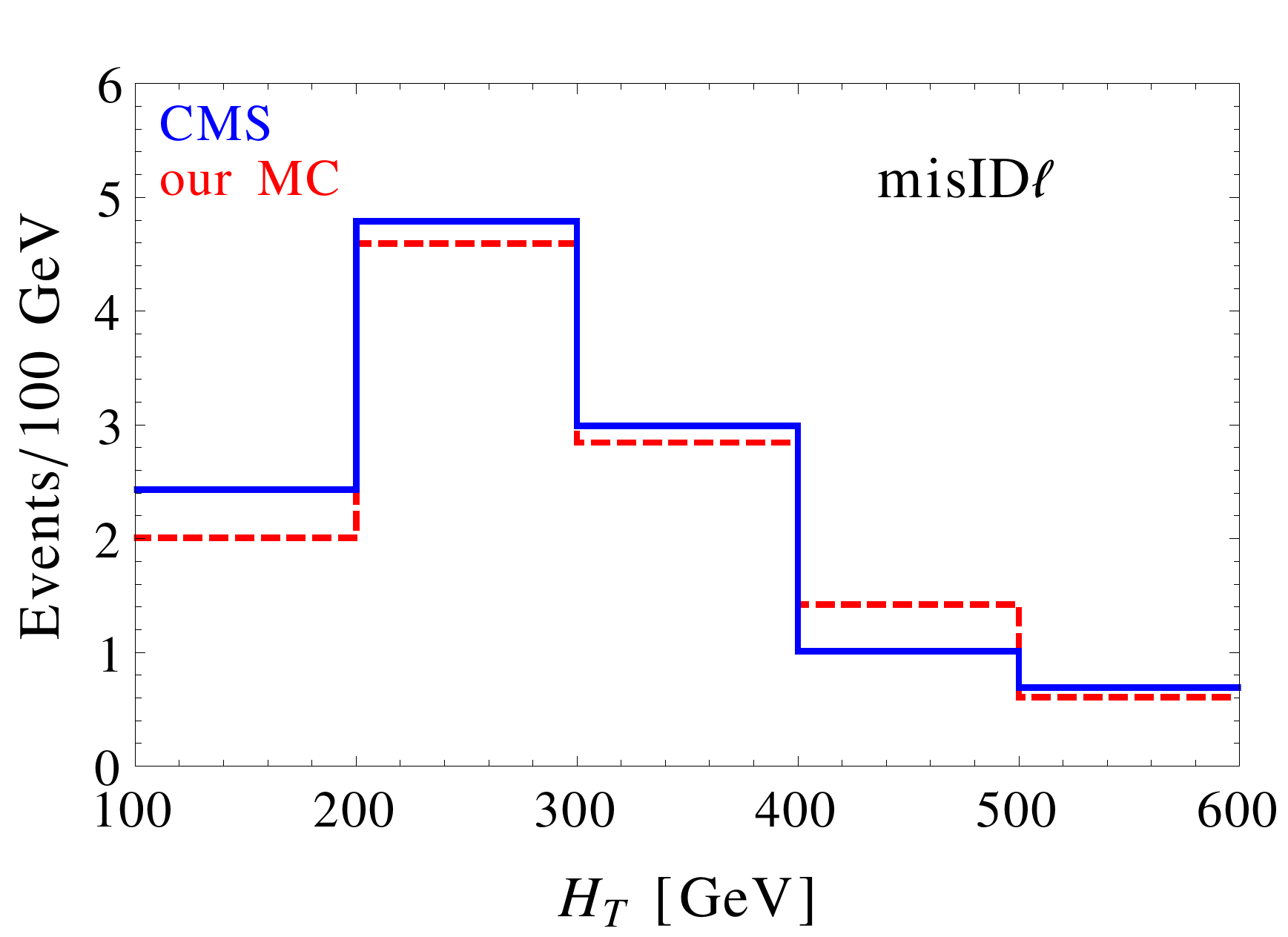}
\includegraphics[width=0.305\textwidth]{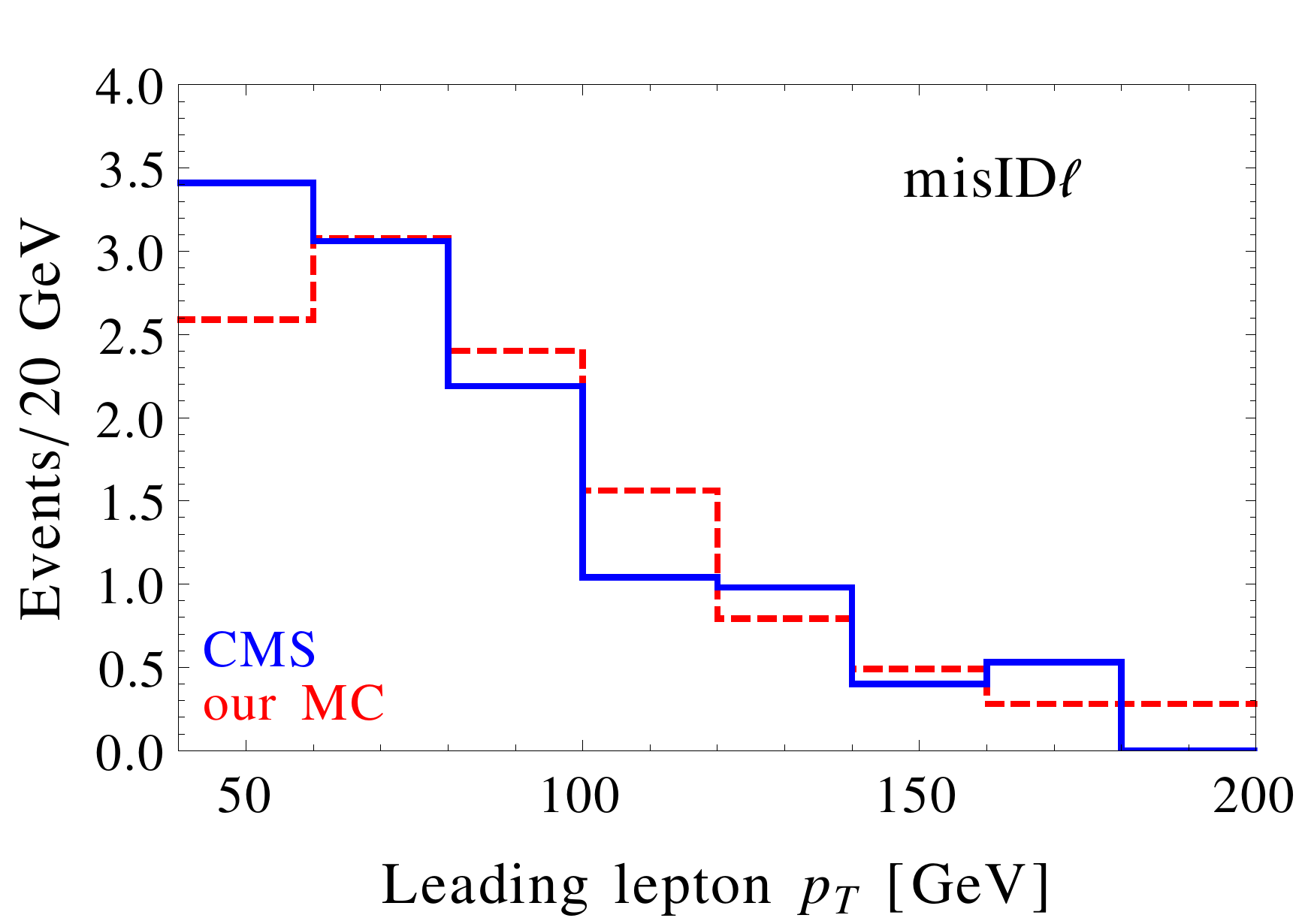}
 \end{center}
 \caption{$8$ TeV distributions for $(t\bar{t}W$+jets$)_{\mathrm{QCD}}$ (upper row) and misID$\ell$ (lower row). The blue histograms show the CMS result, whereas the red histograms show our prediction, after normalizing to the CMS total yields.}
\label{fig:8TeVdistributions}
\end{figure}

To set bounds from the $8$ TeV CMS data, it was sufficient to simulate the $(t\bar{t}Wj)_{\mathrm{EW}}$ signal and make use of the expected background yields quoted by CMS. Thus it was not necessary to simulate all the backgrounds listed in Table~\ref{tab:8TeVyields}. However, because our aim is also to devise an analysis at $13$ TeV, which we will specifically tailor to improve the sensitivity to top-$Z$ interactions, we first need to make sure that our simulation can reproduce the $8$ TeV results contained in Ref.~\cite{CMSttV} for all the processes listed in Table~\ref{tab:8TeVyields}. The salient features of the simulation are summarized below for each background. Unless otherwise specified, jet matching is performed using the shower $k_{\bot}$ scheme \cite{kTshower} with matching scale set to $30$ GeV, and the definition of jet includes $b$-jets.
\begin{itemize}
\item $(t\bar{t}W$+jets$)_{\mathrm{QCD}}$: we generate a matched sample of $t\bar{t}W+0,1$ jets, and normalize it to the NLO cross section of $206$ fb \cite{CMSttV}. Notice that in Ref.~\cite{MaltoniPaganiTsinikos} the $(t\bar{t}Wj)_{\mathrm{QCD}}$ component was shown to have dramatic effects in some regions of phase space. However, the NLO corrections to $(t\bar{t}Wj)_{\mathrm{QCD}}$ were also computed, finding that they have a small effect.\footnote{We thank F.~Maltoni for pointing this out to us.} This supports the use of a LO matched sample with $0,1$ jets, at least for our exploratory analysis. 
\item MisID$\ell$: CMS estimated this background by means of a data-driven method. We follow the approach of Ref.~\cite{FakeLeptons}, where it was proposed to exploit the relationship between the misidentified (or `fake') lepton and the heavy flavor jet from which it originated. The method consists of applying certain probability and transfer functions to MC events containing heavy flavor jets. In our case, we consider a matched sample of $t\bar{t}+0,1,2$ jets, normalized to the NNLO cross section of $245.8\;\mathrm{pb}$ \cite{ttNNLO}. More details about the method are given in App.~\ref{App:FakeLeptons}. Here we only stress that the overall efficiency of the fake lepton generation is a free parameter of the method, and we simply fix it to reproduce the CMS yield reported in Table~\ref{tab:8TeVyields}.   
\item Irreducible: this background is composed mainly by $\bar{t}t$ production in association with a Higgs, with a $\sim 10\%$ component of same-sign $WW$. For the former process, we generate a matched sample of $t\bar{t}h+0,1$ jets, and normalize it to the NLO cross section of $129.3$ fb \cite{lhcxswg}. For the latter, we generate $W^{\pm}W^{\pm}+3j$ with matching, with LO normalization.
\item $t\bar{t}Z$: we generate a matched sample of $t\bar{t}Z+0,1$ jets, and normalize it to the NLO cross section of $197$ fb \cite{CMSttV}.
\item misID$Q$: this background was estimated by CMS using a combination of data and MC. We mimic their method by selecting MC events that contain opposite-sign $e\mu$ or $ee$ and pass all the cuts except for the same-sign requirement, and weighting them with the probability for the charge of each electron to be mismeasured (the probability of the charge of a $\mu$ being mismeasured is negligible). We take the probabilities to be $2.3\times 10^{-3}$ for the endcaps ($|\eta| > 1.479$) and $2\times 10^{-4}$ for the barrel ($|\eta| < 1.479$) \cite{CMSePerformance}. These probabilities correspond to the `selective' charge identification method \cite{CMSePerformance}, and agree with the order-of-magnitude values quoted in Ref.~\cite{CMSttV}. Two processes contribute: $t\bar{t}$, and DY$+\mathrm{jets}$. For the former, which amounts to $\sim 70\%$ of the total, we generate a matched sample of $t\bar{t}+0,1,2$ jets, normalized to the NNLO cross section of $245.8\;\mathrm{pb}$ \cite{ttNNLO}. For the latter, we generate a sample of $\ell^{+}\ell^{-}$+$3j$ with matching, with LO normalization.  
\item $WZ$: we generate $WZ$+$3j$ with matching and LO normalization.
\end{itemize}         
%
\begin{table}[t]
   \begin{center}
   \begin{tabular}{c|c c c c c c} 
   & $(t\bar{t}W$+jets$)_{\mathrm{QCD}}$ & misID$\ell$ & irreducible & $t\bar{t}Z$ & misID$Q$ & $WZ$  \\[0.02cm]
   \hline
   &&&&&& \\[-0.4cm]
   $\frac{\mathrm{CMS}}{\mathrm{our\;MC}}$ & $0.62$ & -- & $1.09$ & $1.20$ & $0.28$ & $0.83$  \\[0.06cm]
   \end{tabular}   
   \end{center}
   \caption{Multiplicative factors we need to apply to the normalization of our MC samples to match the CMS results in Table~\ref{tab:8TeVyields}. The normalization of the misID$\ell$ background is not predicted by the fake lepton simulation.}
   \label{tab:8TeVrescalings}
\end{table}
For each of the above processes, we compare the distributions of the leading lepton $p_T$ and $H_T$ with those reported in Fig.~2 of Ref.~\cite{CMSttV}. The shapes of the distributions are reproduced in all cases, including misID$\ell$ and misID$Q$, which were predicted by CMS using data. This gives us, in particular, confidence in our treatment of the fake leptons, which together with $(t\bar{t}W$+jets$)_{\mathrm{QCD}}$ will dominate the background in the $13$ TeV analysis. On the other hand, as shown in Table~\ref{tab:8TeVrescalings}, the normalization agrees reasonably well with the CMS result for all the processes except misID$Q$, for which our simulation overestimates the event yield by a factor $\sim 3.5$. Nevertheless, once we normalize to the CMS rate, the misID$Q$ distributions are reproduced to good accuracy. In addition, we checked that for all processes we reproduce, within errors, the relative contributions to the $6$ SSL categories shown in Table~1 of Ref.~\cite{CMSttV}. The $8$ TeV distributions for $(t\bar{t}W$+jets$)_{\mathrm{QCD}}$ and misID$\ell$ are shown in Fig.~\ref{fig:8TeVdistributions}, after normalization to the respective CMS yields. Having validated our background simulations against data, we will confidently make use of them in the $13$ TeV analysis, by generating each process with the same settings employed at $8$ TeV, including jet multiplicity. The normalization will be fixed to the best available calculation (see Table~\ref{tab:13TeVxsec}), multiplied by the rescaling factor given in Table~\ref{tab:8TeVrescalings}, which brings our $8$ TeV rate in agreement with the one predicted by CMS.\footnote{The irreducible and misID$Q$ backgrounds are composed by two distinct processes. In these cases, the rescaling factor of Table~\ref{tab:8TeVrescalings} is applied to each of the component processes.} The misID$\ell$ process will be simulated with the same parameters that match the CMS results at $8$ TeV, because we do not expect a significant variation going to the higher collider energy.     
%
%

%
\begin{table}[h]
   \begin{center}
   \begin{tabular}{c|c c c c} 
   & $(t\bar{t}W$+jets$)_{\mathrm{QCD}}$ & $t\bar{t}$ (misID$\ell$, misID$Q$) & $t\bar{t}h$ & $t\bar{t}Z$   \\[0.02cm]
   \hline
   &&&& \\[-0.4cm]
   $\sigma_{\mathrm{13}\;\mathrm{TeV}}$ & $566.3\;\mathrm{fb}$ \cite{ttW13NLO} & $788.2\;\mathrm{pb}$ \cite{Hathor} & $508.5\;\mathrm{fb}$ \cite{LHCxswg} & $771\;\mathrm{fb}$ \cite{ttW13NLO} \\[0.06cm]
   \end{tabular}   
   \end{center}
   \caption{Inclusive cross sections used to normalize the $13$ TeV samples. The $t\bar{t}$ cross section is at approximate NNLO, whereas the others are at NLO. For the processes not listed here, LO normalization was used.}
   \label{tab:13TeVxsec}
\end{table}
%
\subsection{13 TeV analysis}
The $(t\bar{t}Wj)_{\mathrm{EW}}$ process is characterized by the presence of a highly energetic forward jet, which provides a natural handle to separate the signal from the background. In our analysis, we thus select a candidate forward jet, and make use of kinematic variables constructed out of it. However, forward jet tagging is known to face issues in high-pileup conditions, like those of LHC Run-2, and what level of performance will be achieved is still an open question. Interestingly, it was suggested \cite{PerezFWJ}, in the context of a study of heavy top partners in the very $t\bar{t}Wj$ final state, that clustering forward jets with a radius parameter smaller than the standard one can greatly improve the forward jet tagging. Yet in our analysis we go beyond tagging, making use of the reconstructed four-momentum of the forward jet. Because this aspect of the analysis may be affected by pileup, we choose to also perform a separate analysis where we do not make any reference to forward jets, and only employ central jets with $|\eta| < 2.4$. The results of this second analysis (which will be labeled $3j$ analysis) are very robust and likely conservative, whereas the first ($4j$ analysis) illustrates the potential of forward jet variables in suppressing the background.
\subsubsection{$4j$ analysis}\label{4janalysis}
\begin{figure}[t!]
 \begin{center}
\includegraphics[width=0.325\textwidth]{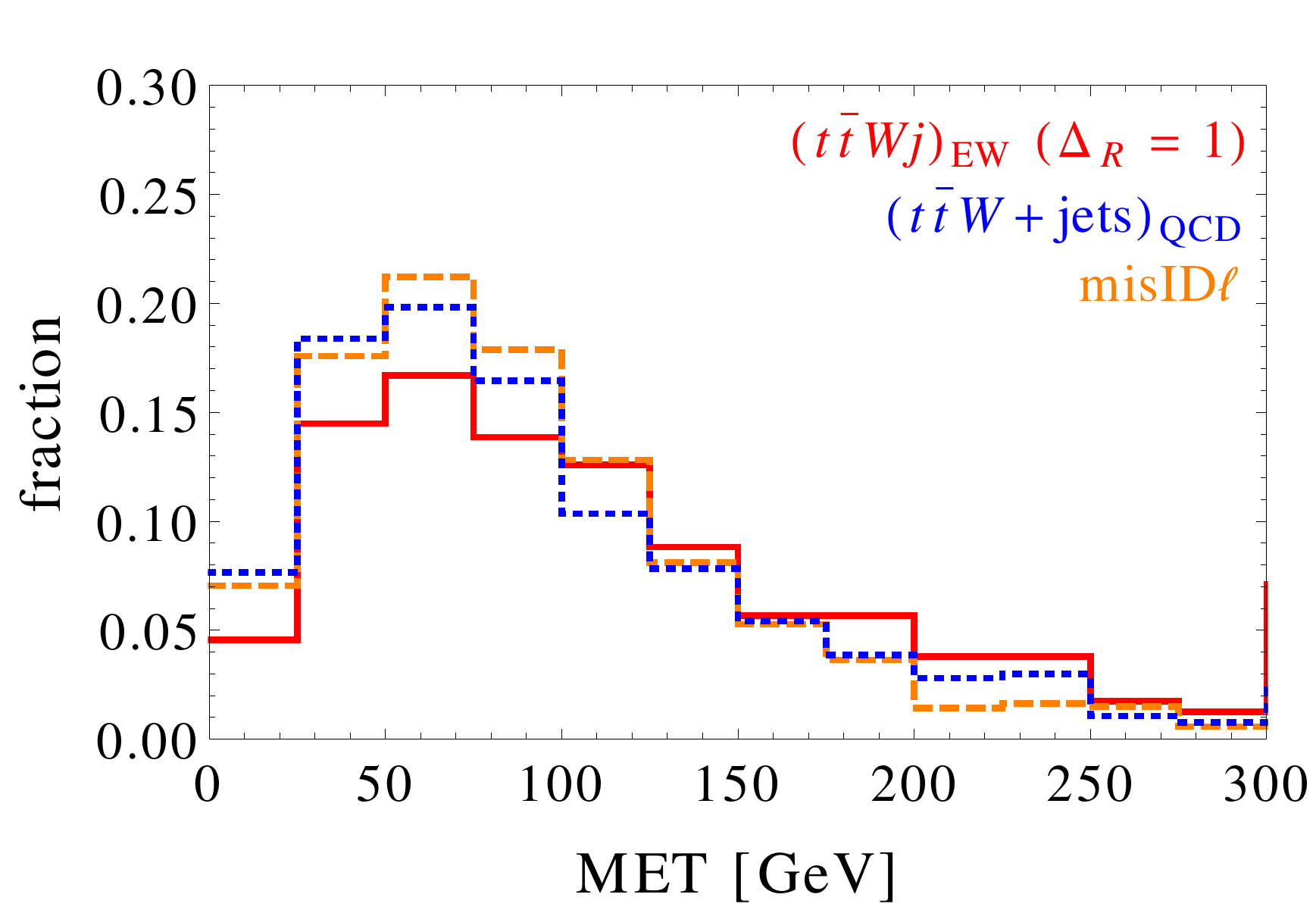}
\includegraphics[width=0.335\textwidth]{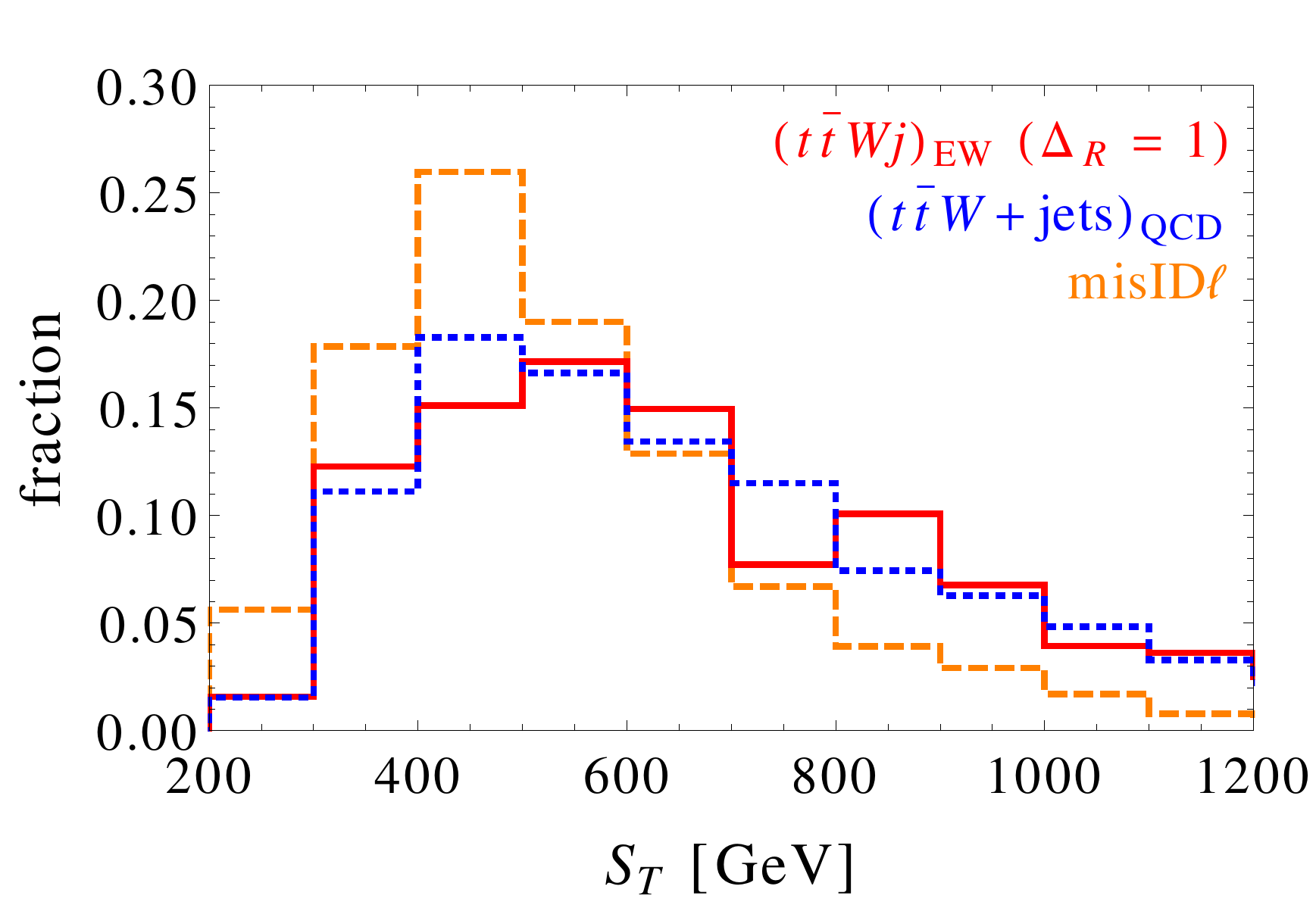}
\includegraphics[width=0.325\textwidth]{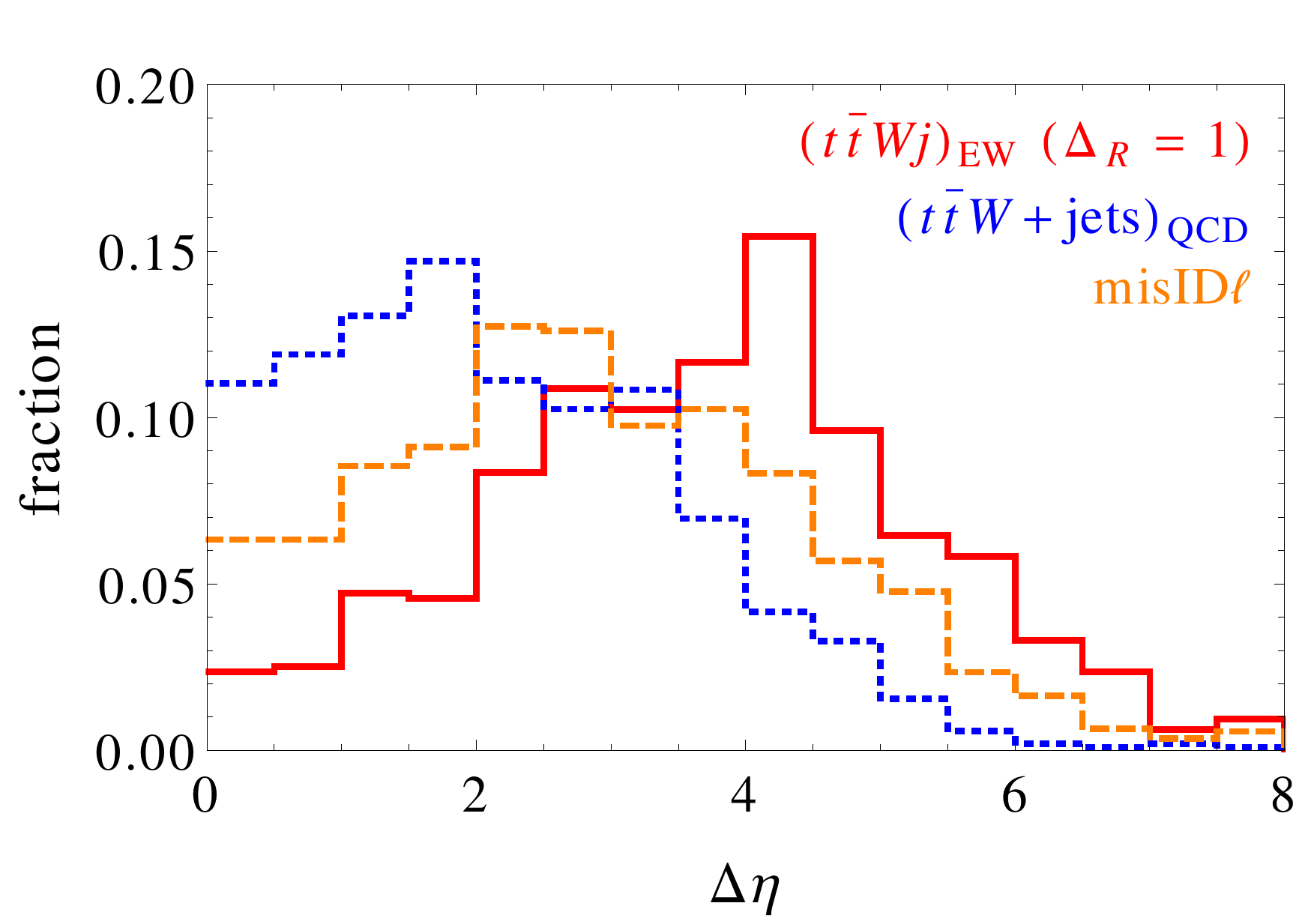}
 \end{center}
 \caption{Normalized distributions for $(t\bar{t}Wj)_{\mathrm{EW}}$ and the two main backgrounds $(t\bar{t}W$+jets$)_{\mathrm{QCD}}$ and misID$\ell$ at $13$ TeV, after the $4j$ pre-selection. The other relevant distributions were shown in Fig.~\ref{fig:13TeVdistribsec3}.}
\label{fig:13TeVdistribsec4}
\end{figure}

In the $4j$ analysis, we make use of the forward jet that characterizes the signal. The event pre-selection requires the following:
\begin{enumerate}
\item The cuts on leptons are identical to the $8$ TeV analysis;
\item We require at least four jets with $p_{T}> 30$ GeV and $|\eta| < 5$, among which at least three must be central, {\it i.e.} must satisfy $|\eta| < 2.4$ (at least one of the central jets must be $b$-tagged), and at least one must not be $b$-tagged;
\item No cut is applied on $H_T$, defined as the scalar sum of the $p_T$ of all the central jets as defined in point 2;
\item The CMS lepton isolation is approximated by requiring that $\Delta R (\ell, j) > 0.3$ for each of the SSL and for all central jets. 
\end{enumerate}
After the pre-selection, to find the best set of cuts we perform an optimization based on the signal point $(\Delta_L,\Delta_R) = (0,1)$, which corresponds roughly to the target sensitivity at $13$ TeV with $300$ fb$^{-1}$. For the optimization, we include only the two main backgrounds $(t\bar{t}W+\mathrm{jets})_{\mathrm{QCD}}$ and misID$\ell$. The optimization is performed by maximizing the statistical significance of the exclusion\footnote{The $95\%$ CL exclusion corresponds to $\mathcal{S} \simeq 1.64$.}
\begin{equation} \label{significance}
\mathcal{S}\equiv \frac{S}{\sqrt{S+B}} = \frac{N_{(t\bar{t}Wj)_{\mathrm{EW}}}(\Delta_R = 1) - N_{(t\bar{t}Wj)_{\mathrm{EW}}}(\mathrm{SM})}{\sqrt{N_{(t\bar{t}Wj)_{\mathrm{EW}}}(\Delta_R = 1) + N_{(t\bar{t}W+\mathrm{jets})_{\mathrm{QCD}}} + N_{\mathrm{misID}\ell} }}\;,
\end{equation}    
where $S$ and $B$ indicate the signal and background, respectively. The luminosity is assumed to be $300\;\mathrm{fb}^{-1}$.
We consider a number of candidate variables in order to enhance $\mathcal{S}$. The best are found to be the transverse momentum of the leading lepton, $p_T^{\ell_1}$, the invariant mass of the two leading leptons, $m_{\ell_1 \ell_2}$, the missing transverse energy, $\mathrm{MET}$, the scalar sum of $H_T$ and the $p_T$ of the two leading leptons, $S_T$, and two angular variables that involve the forward jet, namely $|\eta_{j_{\mathrm{fw}}}|$ and $\Delta\eta \equiv |\eta_{j_{\mathrm{fw}2}}-\eta_{j_{\mathrm{fw}}}|\,$, where the forward jet $j_{\mathrm{fw}}$ and the `second forward jet'  $j_{\mathrm{fw}2}$ are defined as
\begin{itemize}
\item $j_{\mathrm{fw}}$ is the non-$b$-tagged jet with largest $|\eta|\,$,
\item $j_{\mathrm{fw}2}$ is the jet with the largest invariant mass with $j_{\mathrm{fw}}$.
\end{itemize}
The normalized signal and background distributions of these variables after the pre-selection cuts are shown in Figs.~\ref{fig:13TeVdistribsec3} and \ref{fig:13TeVdistribsec4}. The cut-flow for our optimal cuts is in Table~\ref{tab:cutflow13_4j}.
%
%
%
\begin{table}[h!]
   \begin{center}
   \begin{tabular}{c|c|c c c c|c} 
    & $\mathcal{S}$ & EW$(\mathrm{SM})$ & EW$(\Delta_R = 1)$ & $(t\bar{t}W$+jets$)_{\mathrm{QCD}}$ & misID$\ell$ & $S/B$ \\[0.01cm]
   \hline
   &&&&&& \\[-0.45cm]
   pre-selection  & $2.8$ & $91$ & $183$ & $445$ & $476$ & $0.091$\\[0.01cm]
   \hline
      &&&&&& \\[-0.45cm]    
   $p_T^{\ell_1} > 100$ GeV & $3.0$ & $44$ & $111$ & $223$ & $166$ & $0.15$ \\[0.01cm]     
   \hline
      &&&&&& \\[-0.45cm]
   $m_{\ell_1\ell_2} > 125$ GeV & $3.1$ & $39$ & $102$ & $202$ & $112$ & $0.18$ \\[0.01cm]
   \hline
      &&&&&& \\[-0.45cm]
   $\mathrm{MET} > 50$ GeV & $3.2$ & $28$ & $84$ & $152$ & $80$ & $0.22$ \\[0.01cm]
   \hline
       &&&&&& \\[-0.45cm]
   $|\eta_{j_{\mathrm{fw}}}| > 1.75$ & $3.4$ & $21$ & $69$ & $77$ & $58$ & $0.31$ \\[0.01cm]
   \hline
      &&&&&& \\[-0.45cm]
   $\Delta\eta > 2$ & $3.5$ & $20$ & $67$ & $60$ & $49$ & $0.36$ \\[0.01cm]
   \hline
         &&&&&& \\[-0.45cm]
   $S_T > 500$ GeV & $3.5$ & $16$ & $58$ & $51$ & $34$ & $0.42$ \\[0.01cm]
   \end{tabular}   
   \end{center}
   \caption{Cut-flow for the $4j$ optimization at $13$ TeV. EW stands for $(t\bar{t}Wj)_{\mathrm{EW}}$.}
   \label{tab:cutflow13_4j}
\end{table}
We see that the cuts on the leptons effectively suppress the fake lepton background, while the cuts on the forward jet are successful against the $(t\bar{t}W$+jets$)_{\mathrm{QCD}}$ background. After all cuts, we achieve a significance of $3.5$ and a signal to background ratio of approximately $0.4$. In Table~\ref{tab:13TeVsubleadingYields4j} we report the event yields for the subleading backgrounds. We note that because our selection requires at least $4$ jets, the backgrounds $W^{\pm}W^\pm, WZ$ and DY should in principle be simulated with four extra partons in the matrix element, matched to the parton shower. This requires, however, a large computational effort, which goes beyond the scope of this paper. Therefore, as an estimate, we simply simulate these backgrounds with $3$ additional partons at matrix element level. We find that all three processes give a very small contribution to the signal region. In particular, DY+jets is very strongly suppressed by the MET cut.   
%
\begin{table}[h]
   \begin{center}
   \begin{tabular}{c|c c c c c c} 
   & $t\bar{t}h$ & $W^{\pm}W^{\pm}$ & $t\bar{t}Z$ & $t\bar{t}$-misID$Q$ & DY-misID$Q$ & $WZ$   \\[0.02cm]
   \hline
   &&&& \\[-0.4cm]
   pre-selection & $233$ & $18$ & $105$ & $44$ & $16$ & $41$ \\[0.06cm]
   \hline
      &&&& \\[-0.4cm]
   all cuts & $19$ & $3$ & $8$ & $4$ & $0$ & $4$ \\[0.06cm]
   \end{tabular}   
   \end{center}
   \caption{Event yields for the subleading backgrounds at $13$ TeV, after $4j$ pre-selection and after the full $4j$ analysis.}
   \label{tab:13TeVsubleadingYields4j}
\end{table}
The signal yields after all cuts were given in Eqs.~(\ref{13TeVyieldcoupl4j},~\ref{13TeVyieldhdo4j}).
The statistical uncertainty on the signal yields computed using those equations is approximately $10\%$. To derive the constraints on the parameters $\vec{p}$, we perform a single-bin\footnote{We have verified that taking one inclusive bin instead of $6$ SSL categories makes the $8$ TeV bounds only slightly weaker.} likelihood analysis, in complete analogy with Eq.~\eqref{likelihood}, assuming the observed number of events to equal the SM prediction. We also consistently take into account the dependence of the subleading background $t\bar{t}Z$ on the parameters $\vec{p}$, by rescaling the yield in Table~\ref{tab:13TeVsubleadingYields4j} with the ratio $\sigma_{t\bar{t}Z}(\vec{p}\,)/\sigma_{t\bar{t}Z}^\mathrm{SM}$, with $\sigma_{t\bar{t}Z}$ the inclusive cross section for $pp\to t\bar{t}Z$ at $13$ TeV. The resulting bounds were shown as red contours in Fig.~\ref{fig:13TeVbound_4j}. In addition to the $20\%$ systematic uncertainty on the signal, we assume either no systematics on the background (solid contours), or $50\%$ systematic uncertainty on the misID$\ell$ component (dotted). For comparison, we also show in dashed blue the results of the $t\bar{t}Z$ projection made in Ref.~\cite{RS}. Assuming that the only deformation of the SM is a modification of the $Z t_R t_R$ coupling, we obtain the one-dimensional bounds reported in Eq.~\eqref{ZtRtRonedim13}.

\subsubsection{$3j$ analysis}
In the $3j$ analysis, we conservatively do not make use of the forward jet that characterizes the signal, and only impose selection cuts on central jets. The pre-selection is identical to the $8$ TeV analysis, except that no requirement on $H_T$ is applied. As in the $4j$ analysis, after the pre-selection we perform a cut optimization taking the point $(\Delta_L, \Delta_R) = (0,1)$ as signal benchmark, using the statistical significance defined in Eq.~\eqref{significance}. The most effective variables in enhancing the significance are found to be $p_T^{\ell_1},m_{\ell_1\ell_2}$, the MET and $S_T$. The cut-flow for the optimal cuts is in Table~\ref{tab:cutflow13_3j}.    
\begin{table}[h!]
   \begin{center}
   \begin{tabular}{c|c|c c c c|c} 
    & $\mathcal{S}$ & EW$(\mathrm{SM})$ & EW$(\Delta_R = 1)$ & $(t\bar{t}W$+jets$)_{\mathrm{QCD}}$ & misID$\ell$ & $S/B$ \\[0.01cm]
   \hline
   &&&&&& \\[-0.45cm]
   pre-selection  & $2.5$ & $108$ & $212$ & $678$ & $788$ & $0.066$\\[0.01cm]
   \hline
      &&&&&& \\[-0.45cm]    
   $p_T^{\ell_1} > 100$ GeV & $2.9$ & $52$ & $129$ & $346$ & $258$ & $0.12$ \\[0.01cm]     
   \hline
      &&&&&& \\[-0.45cm]
   $m_{\ell_1\ell_2} > 125$ GeV & $2.9$ & $45$ & $117$ & $308$ & $170$ & $0.14$ \\[0.01cm]
   \hline
      &&&&&& \\[-0.45cm]
   $\mathrm{MET} > 50$ GeV & $3.0$ & $32$ & $96$ & $229$ & $122$ & $0.17$ \\[0.01cm]
   \hline
      &&&&&& \\[-0.45cm]
   $S_T > 500$ GeV & $3.0$ & $25$ & $82$ & $186$ & $80$ & $0.19$ \\[0.01cm]
   \end{tabular}   
   \end{center}
   \caption{Cut-flow for the $3j$ optimization at $13$ TeV. EW stands for $(t\bar{t}Wj)_{\mathrm{EW}}$.}
   \label{tab:cutflow13_3j}
\end{table}
After all cuts, we find a significance of $3.0$ and a signal to background ratio of $\sim 0.2$, to be compared with $3.5$ and $\sim 0.4$, respectively, for the $4j$ analysis. The event yields for the subleading backgrounds are reported in Table~\ref{tab:13TeVsubleadingYields3j}. The DY-mis$Q$ background is very strongly suppressed by the MET cut. 
%
\begin{table}[t]
   \begin{center}
   \begin{tabular}{c|c c c c c c} 
   & $t\bar{t}h$ & $W^{\pm}W^{\pm}$ & $t\bar{t}Z$ & $t\bar{t}$-misID$Q$ & DY-misID$Q$ & $WZ$   \\[0.02cm]
   \hline
   &&&& \\[-0.4cm]
   pre-selection & $324$ & $32$ & $188$ & $81$ & $32$ & $62$ \\[0.06cm]
   \hline
      &&&& \\[-0.4cm]
   all cuts & $35$ & $12$ & $36$ & $10$ & $0$ & $16$ \\[0.06cm]
   \end{tabular}   
   \end{center}
   \caption{Event yields for the subleading backgrounds at $13$ TeV, after $3j$ pre-selection and after the full $3j$ analysis.}
   \label{tab:13TeVsubleadingYields3j}
\end{table}
The signal yields after all cuts are found to be
\begin{align} \label{13TeVyieldcoup3j}
N_{(t\bar{t}Wj)_{\mathrm{EW}}}(\Delta_L, \Delta_R) \,=&\,\,27.1 + 21.1\, \Delta_L + 240.2\, \Delta_L^2 + 3.2\, \Delta_R + 50.4 \,\Delta_L \Delta_R + 58.0\, \Delta_R^2 \\\label{13TeVyieldhdo3j}
N_{(t\bar{t}Wj)_{\mathrm{EW}}} (\cL, \cR) \,=&\,\, 27.1 - 98.4\, \cL + 190.4\, \cL^{\,2} - 251.2\, \cL^{\,3} + 597.2\, \cL^{\,4} + 6.5\, \cR \nonumber\\\,&\,\qquad\qquad\qquad\qquad\qquad\;\;\; - 
 0.5\, \cL \cR - 202.6\, \cL^{\,2} \cR + 
 591.2\, \cR^{\,2} \,,
\end{align} 
The statistical uncertainty on the signal yields computed using Eqs.~\eqref{13TeVyieldcoup3j} and \eqref{13TeVyieldhdo3j} is approximately $10\%$.
%
\begin{figure}[t!]
 \begin{center}
\includegraphics[width=0.365\textwidth]{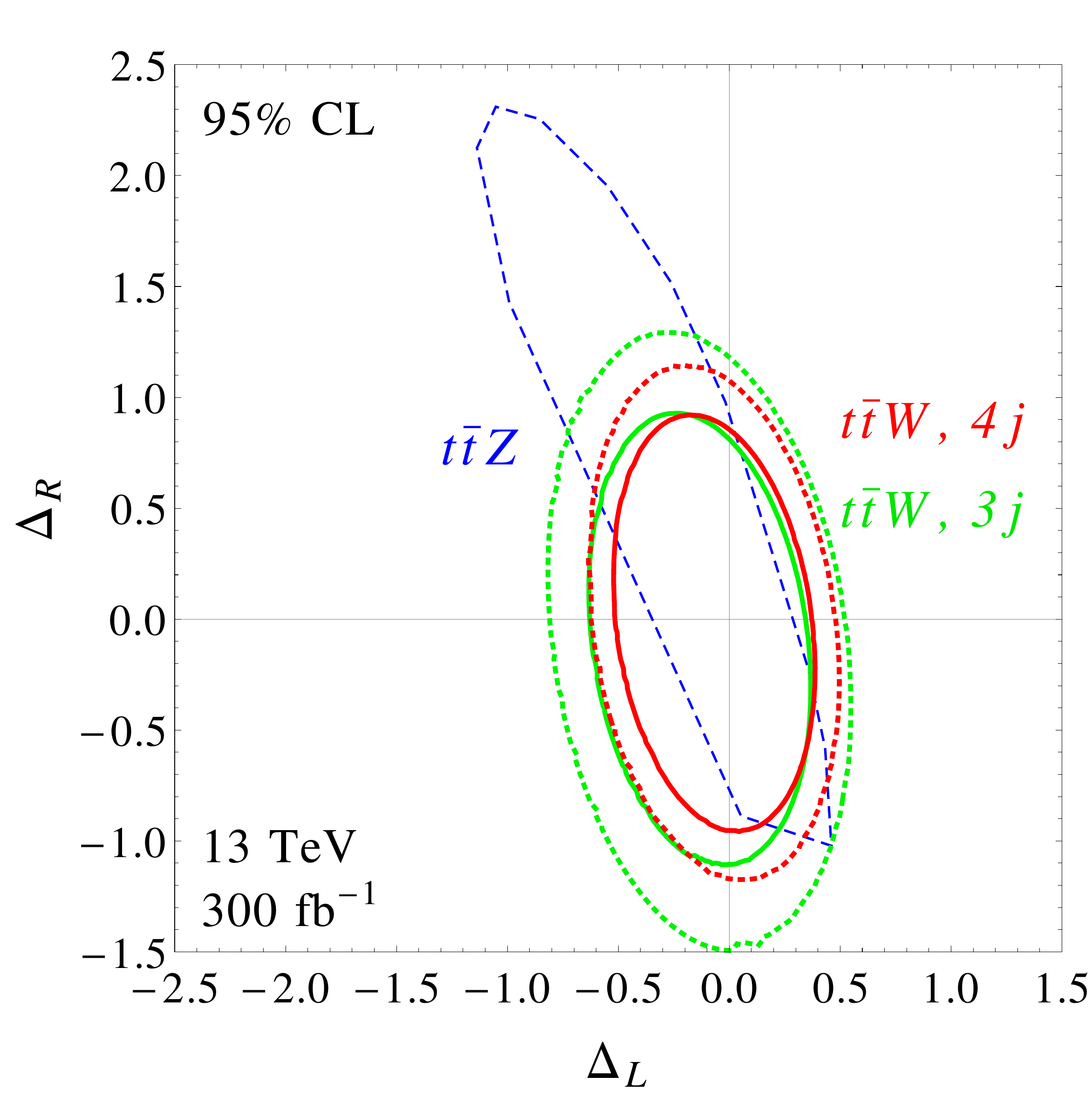}
\includegraphics[width=0.36\textwidth]{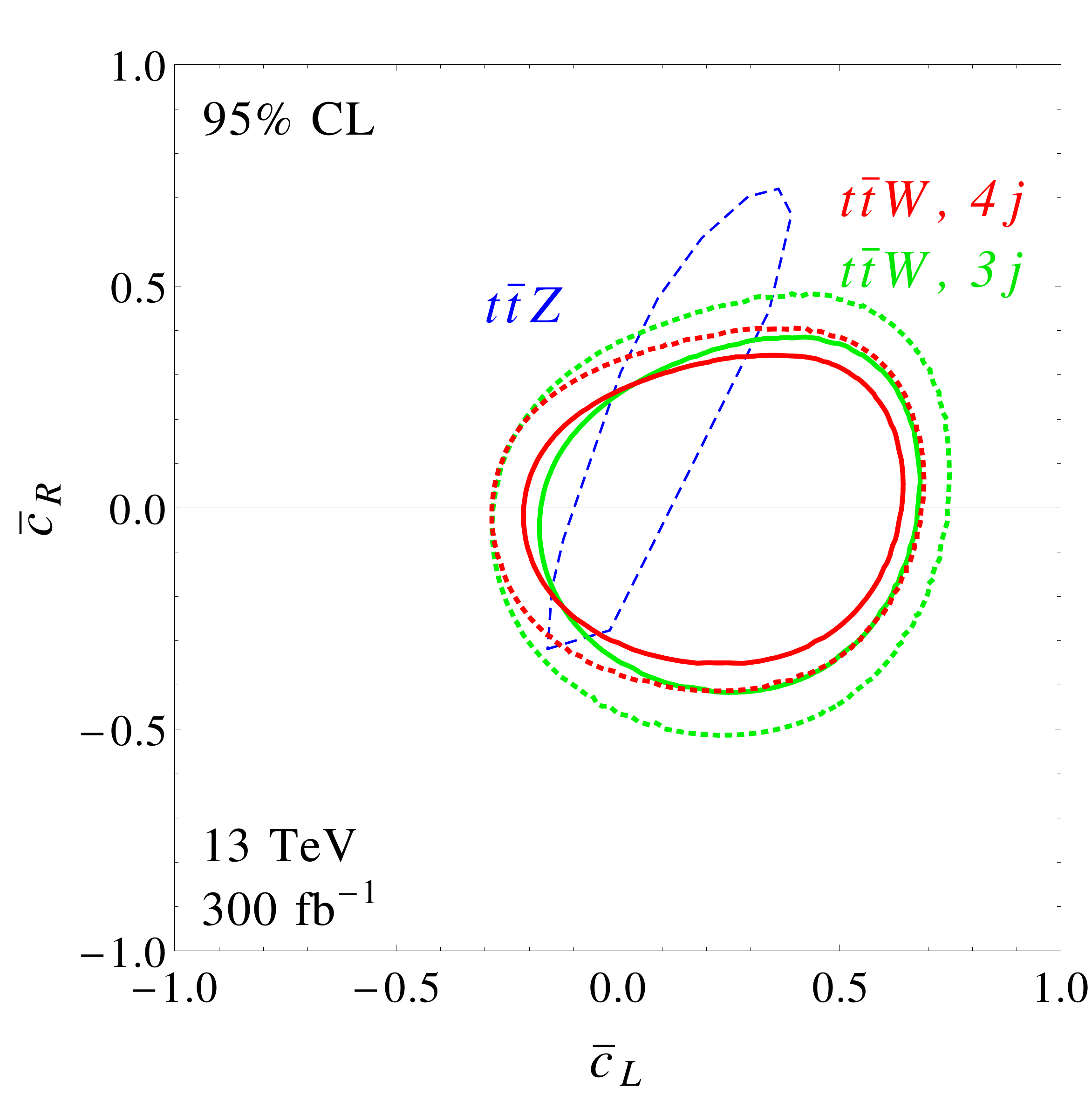}\\
 \end{center}
 \caption{In green, the constraints on top-$Z$ coupling deviations (left panel) and HDO coefficients (right panel) derived from our $3j$ $t\bar{t}W$ analysis at $13$ TeV. The solid contour assumes no systematic uncertainty on the background, whereas the dotted one includes a $50\%$ systematic on the misID$\ell$ component. For comparison, in red we show the corresponding constraints derived from the $4j$ $t\bar{t}W$ analysis, and in dashed blue the constraint obtained from the $13$ TeV $t\bar{t}Z$ analysis, as derived in Ref.~\cite{RS}. The red and blue contours are identical to Fig.~\ref{fig:13TeVbound_4j}.}
\label{fig:13TeVbound_3j}
\end{figure}
%
To set constraints on the parameters $\vec{p}$ we follow exactly the same procedure described for the $4j$ analysis, including taking into account the contamination due to the $t\bar{t}Z$ process. The resulting bounds are shown as green contours in Fig.~\ref{fig:13TeVbound_3j}.\footnote{From Fig.~\ref{fig:13TeVbound_3j} we read that, in the absence of systematics on the background, the $3j$ analysis gives a stronger constraint than the $4j$ one in the $\Delta_L=0\,$, $\Delta_R > 0$ direction. This may be surprising, considering that we chose this very direction for the optimization of the cuts, and that the $4j$ analysis reached a higher significance ($3.5$ versus $3.0$). The effect is due to the $t\bar{t}Z$ contamination, which slightly shifts all contours, and does so more markedly for the $3j$ analysis, where the $t\bar{t}Z$ contribution to the signal region is larger.} If we assume that the only deformation of the SM is a modification of the $Z t_R t_R$ coupling, we find at $95\%$ CL
\begin{align}
\qquad-0.98 < \Delta_R < 0.70\qquad\, &\mathrm{or}\qquad -0.31 < \cR < 0.22\,\qquad (\mathrm{no\; syst\;on}\; B), \nonumber\\
\qquad-1.34 < \Delta_R < 1.05\qquad\, &\mathrm{or}\qquad -0.42 < \cR < 0.33\,\qquad (\mathrm{50\%\; syst\;on}\; \mathrm{misID}\ell).
\end{align}   
We see that in the $3j$ analysis, the deterioration of the bound due to the large systematic uncertainty on the misID$\ell$ background is stronger than in the $4j$ analysis, where this background is more effectively suppressed by the cuts.  

\subsection{Perturbative unitarity of the hard scattering process}
\begin{figure}[t!]
 \begin{center}
\includegraphics[width=0.45\textwidth]{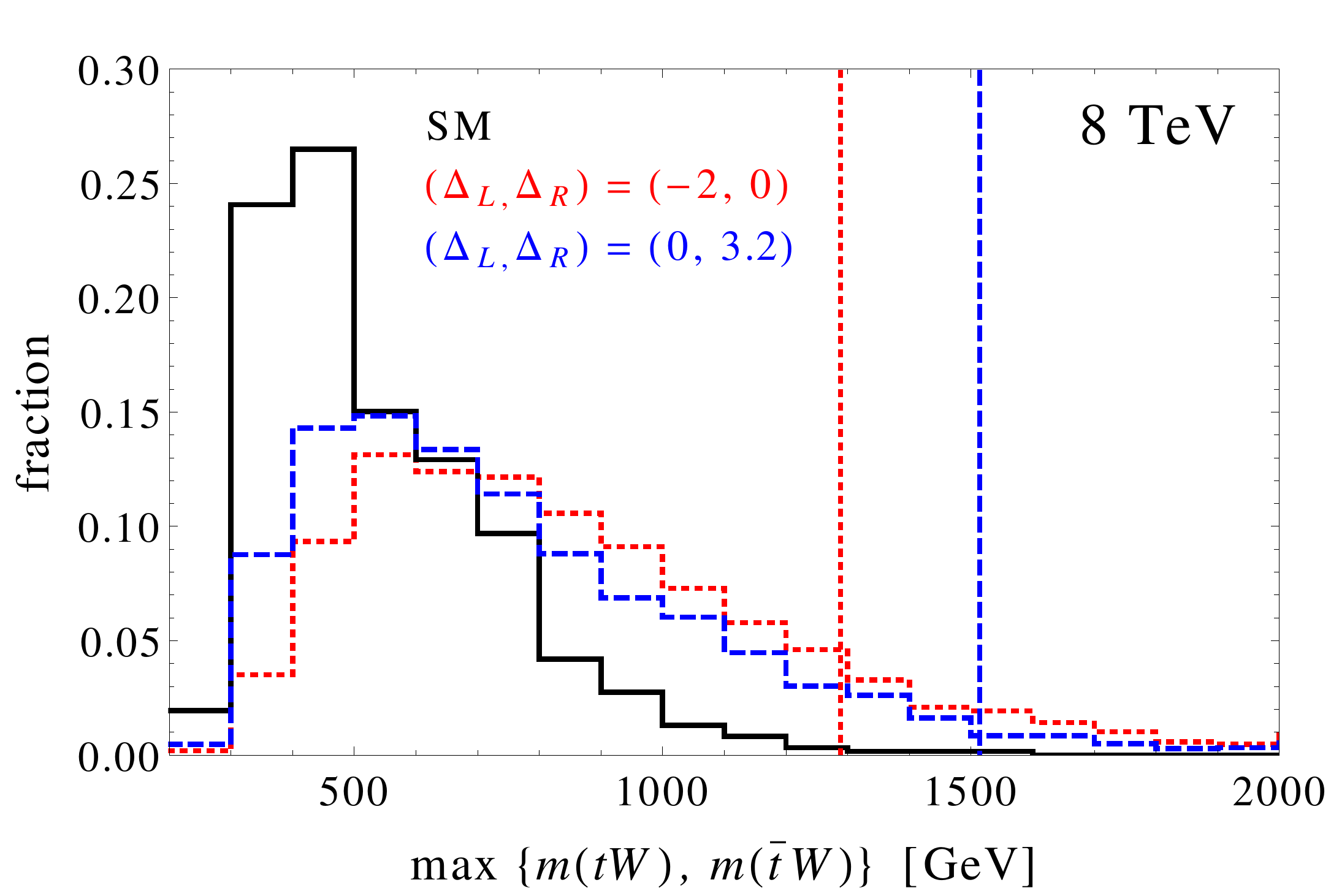}
\includegraphics[width=0.45\textwidth]{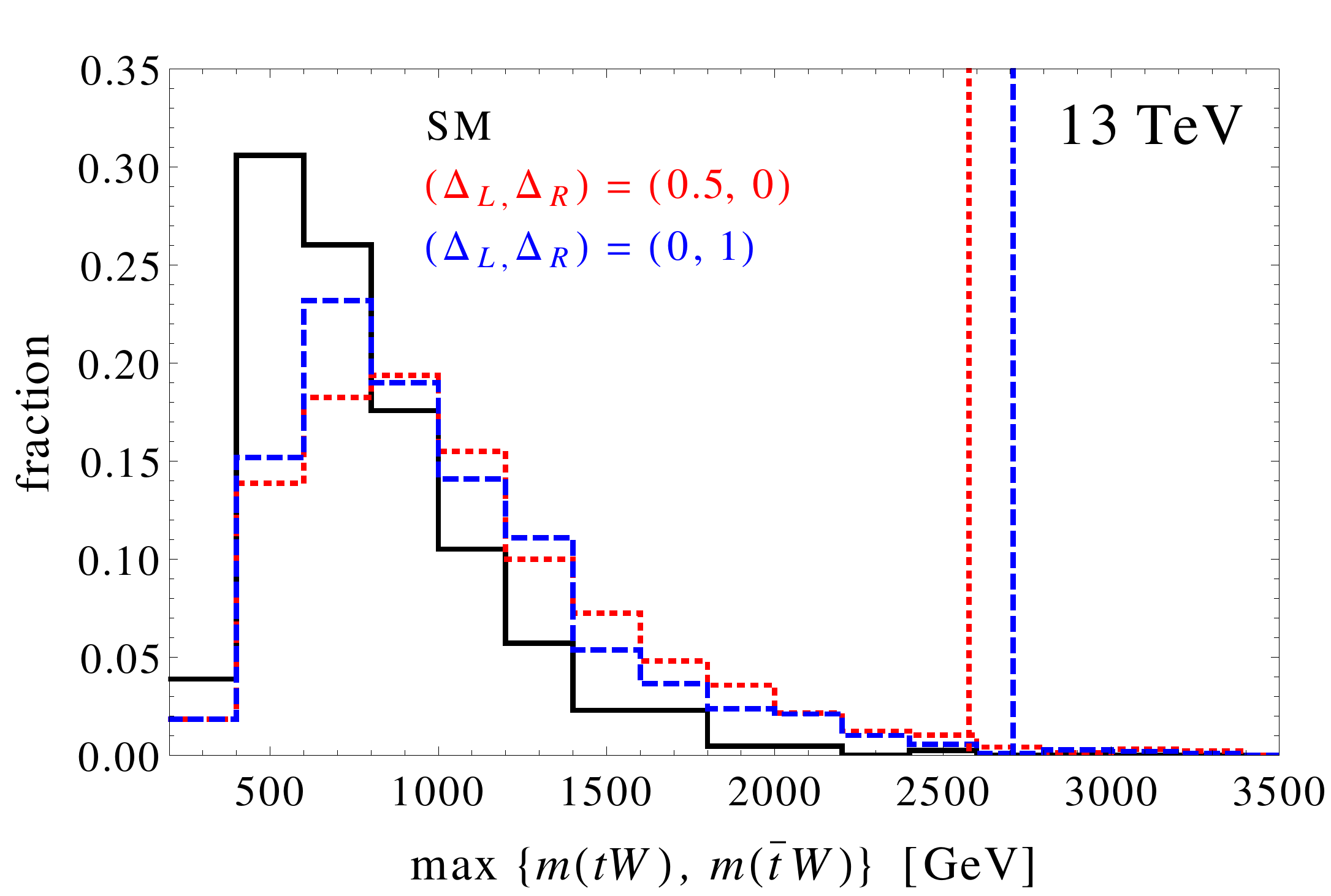}\\
 \end{center}
 \caption{Distributions of the partonic center of mass energy, defined as the largest between $m(tW)$ and $m(\bar{t}W)$, for $(t\bar{t}Wj)_{\mathrm{EW}}$ signal events at $8$ TeV (left panel) and $13$ TeV (right panel). The distributions, shown for a set of representative signal points and for the SM, are obtained after application of all selection cuts. The cutoff scales corresponding to each signal point are also shown as vertical lines.}
\label{fig:unitarity}
\end{figure}

As we discussed at length, the growth with the square of the energy of the $tW \to tW$ scattering amplitude is the reason behind the sensitivity of our analysis to anomalous top-$Z$ couplings. However, this growth also implies that at sufficiently high energy, the amplitude becomes so large that perturbative unitarity is lost, making our predictions not trustable. The scale at which this takes place can be estimated, for example, by computing the $s$-wave amplitude
\begin{equation}
a_0 = \frac{1}{16\pi s}\int_{-s}^0 dt \mathcal{M}\,,
\end{equation}
where $\mathcal{M}$ is the amplitude, and requiring that $|a_0|<1$.\footnote{With this definition, perturbative unitarity in $WW$ scattering is lost, in the absence of a Higgs boson, at the scale $\Lambda = 4\sqrt{2\pi}\,v\simeq 2.5$ TeV.} Considering first deviations in the $Z t_R t_R$ coupling, by integration of the $\mathcal{M}_{RR}$ amplitude in Eqs.~(\ref{GeneralAmplitude}, \ref{tW-tW-}), one finds the following estimate for the cutoff scale
\begin{equation}\label{cutoff-R}
\Lambda = \frac{3\sqrt{\pi}\,v}{s_w\sqrt{|\Delta_R|}}\,,
\end{equation}
which equals $2.7$ TeV for $|\Delta_R| = 1$, corresponding to a BSM contribution of the same size of the SM coupling. Similarly, for deviations in the $Z t_L t_L$ coupling the relevant amplitude is $\mathcal{M}_{LL}$, leading to
\begin{equation}\label{cutoff-L}
\Lambda = \frac{2\sqrt{3\pi}\,v}{\sqrt{1-\frac{4}{3}s_w^2}\,\sqrt{|\Delta_L|}}\,,
\end{equation}
which equals $1.8$ TeV for $|\Delta_L| = 1$. To understand whether perturbative unitarity is an issue in our signal predictions, we should consider the distribution of the center-of-mass energy $\sqrt{\hat{s}}$ of the partonic hard scattering $tW \to tW$ in LHC events. However, given the topology of the signal process $pp\to t\bar{t}Wj$, it is impossible to tell on an event-by-event basis whether the hard scattering that took place was $tW\to tW$, or rather $\bar{t}W\to \bar{t}W$. Thus, to be conservative, for each event we identify $\sqrt{\hat{s}}$ with the largest of the partonic invariant masses $m(tW)$ and $m(\bar{t}W)$. Normalized distributions of this quantity are shown in Fig.~\ref{fig:unitarity}. For each collider energy we show the distributions, obtained after all selection cuts, for a set of signal points that sit approximately at the edge of the exclusion region, together with the corresponding cutoff scales obtained from Eqs.~(\ref{cutoff-R}) and (\ref{cutoff-L}). We observe that even for the very large deviations allowed by $8$ TeV data, the fraction of events whose $\sqrt{\hat{s}}$ could potentially be larger than the cutoff is at most $10\%$. At $13$ TeV, this fraction is approximately $1\%$. We conclude that our predictions are robustly safe from issues with perturbative unitarity.  
%
%

\section{Other processes}\label{sec:OtherProcesses}
In this section we wish to discuss other scattering processes beyond $tW\to tW$ that involve third generation fermions and $W,Z$ or $h$, where BSM deviations lead to a growth with energy. We will focus on the phenomenologically most relevant amplitudes, pointing out the deformations of the SM to which they are most sensitive to, as well as the most promising collider processes where they could be probed. While our formulas for the amplitudes are expressed in terms of general coupling deviations, in the discussion we assume that departures from the SM can be parameterized in terms of dim-6 operators, including those proportional to $-\ctreL = \cunoL \equiv  \cL$, $\cR$, and $\cu$, and neglecting the remaining ones in Eq.~\eqref{Lsilh}. The relation between the HDO coefficients and the coupling deviations can be found in Eqs.~(\ref{Lcoup}--\ref{deltag}). For all processes of the type $\psi_1 + \phi_1 \to \psi_2 + \phi_2$, with $\psi_{1,2}= \{t,b\}$ and $\phi_{1,2}$ the longitudinal $W^{\pm}, Z$ or $h$, we make reference to the general form of the amplitude in Eq.~\eqref{GeneralAmplitude}.
\subsection{$t Z \to t h$}
For $ t Z \rightarrow t h $ we find $\kappa = 1$ and
\begin{align}
A_{LL} &\,=\, \cLth \left(1 - \tfrac{4}{3} s_w^2 \right),\nonumber  \\
A_{RR} &\,=\, - \cRth \tfrac{4}{3} s_w^2\,,\nonumber  \\
A_{LR} &\,=\,  \cLth + \tfrac{1}{2} \cLt ( \ct - \cV ) - \tfrac{2}{3}s_w^2 \left[ 2 \cLth + (\cLt - \cRt) (\ct - \cV) \right],\nonumber \\
A_{RL} &\,=\, - \tfrac{1}{2} \cLt ( \ct - \cV ) - \tfrac{2}{3}s_w^2 \left[ 2 \cRth - (\cLt - \cRt) (\ct - \cV) \right].
\end{align}
This process can be probed in $pp\to t\bar{t}hj$. The leading terms of the amplitude grow as $\hat{s}$ and are controlled by the $hZtt$ interaction, which under our assumptions receives contributions from both $\bar{c}_{L}$ and $\bar{c}_{R}$: we have $A_{LL}\sim \cL$ and $A_{RR} \sim \cR$. As a consequence, $tZ \to th$ can be seen as complementary to $tW\to tW$ in probing these two operators, and in particular $\cR$. However, one important difference between the $ t \bar{t} W j $ and $ t \bar{t} h j $ processes is that for the former the $(t\bar{t}W$+jets)$_{\mathrm{QCD}}$ background is robustly insensitive to new physics in the top sector, whereas for the latter the main background is given by $t\bar{t}h$+jets production, which depends strongly on $\ct = 1 - \cu$. The $t\bar{t}h$ signal has been searched for both by ATLAS~\cite{ATLAS_ttH_bb,ATLAS_ttH_gamma,ATLAS_ttH_multilepton} and CMS~\cite{CMS_ttH}, with $ 8 \text{ TeV} $ data implying an upper limit on the cross section of about $ 3 $ times the SM prediction. To enhance the sensitivity to $\cL$ and $\cR$, one may add to the existing experimental strategy the requirement of a forward jet, as well as additional high energy cuts on the decay products of the Higgs and the tops. It is interesting to note that if an excess were found in $t \bar{t}h$+jets, in principle this could be caused either by $c_t>1$, or by a large deviation of the $Z t_R t_R$ coupling.
\subsection{$b W \to t h$}
For $ b W ^+ \rightarrow t h $ we find $\kappa = \sqrt{2}$ and
\begin{align}
A_{LL} &\,=\,  \cLLh\,, \nonumber \\
A_{RR} &\,=\,  \cRRh\,, \nonumber \\
A_{LR} &\,=\, \cLLh +  \tfrac{1}{2} \cLL ( \ct - \cV ), \nonumber\\
A_{RL} &\,=\, \cRRh + \tfrac{1}{2} \cRR ( \ct - \cV ). \label{ampbWth}
\end{align}
This process can be probed in $pp\to thj$. The leading terms of the amplitude grow as $\hat{s}$ and are controlled by the $hWtb$ interaction, which under our assumptions is generated only by the operator proportional to $\cL$: we have $A_{LL}= - \cL$ and $A_{RR}=0$, thus the leading sensitivity is to $\cL$. An interesting feature of the $bW\to th$ process is that the SM amplitude is strongly suppressed, due to an accidental cancellation between the diagrams with $s$-channel top exchange and $t$-channel $W$ exchange \cite{Maltonith}. As pointed out in Refs.~\cite{Biswasth1,Farinath} (see also Ref.~\cite{Biswasth2}), if only Higgs coupling deviations are considered, this cancellation leads to a striking sensitivity of the cross section to $A_{LR}\sim (c_t - c_V)$, which can be exploited to constrain the sign of $c_t$ through a measurement of the $thj$ process. Following this proposal, the CMS collaboration has performed a full analysis on $8$ TeV data \cite{CMSthcombined}, considering the Higgs decays into $b\bar{b}$, multileptons and $\gamma\gamma$, whereas the ATLAS collaboration has published an analysis in the diphoton channel\cite{ATLAS_ttH_gamma}. We stress that the very strong sensitivity of $thj$ to $(c_t- c_V)$ is mainly due to the threshold region, because of the already mentioned accidental cancellation, thus justifying why the cross section increases by more than one order of magnitude for $c_t = - c_V = -1$, even though the amplitude only grows as $\sqrt{\hat{s}}$ \cite{Farinath}. For more details on the experimental strategy to separate the $thj$ signal from the background, we refer the reader to the analyses in Ref.\cite{CMSthcombined}. Here we simply observe that the growth of the amplitude like $\hat{s}$ in the presence of a non-vanishing $\cL$ suggests the application of tighter cuts on the Higgs and top decay products. In summary, $thj$ may provide an interesting opportunity to constrain $\cL$.

\subsection{$b W \to t Z$}
\begin{figure}[t!]
 \begin{center}
\includegraphics[width=0.6\textwidth]{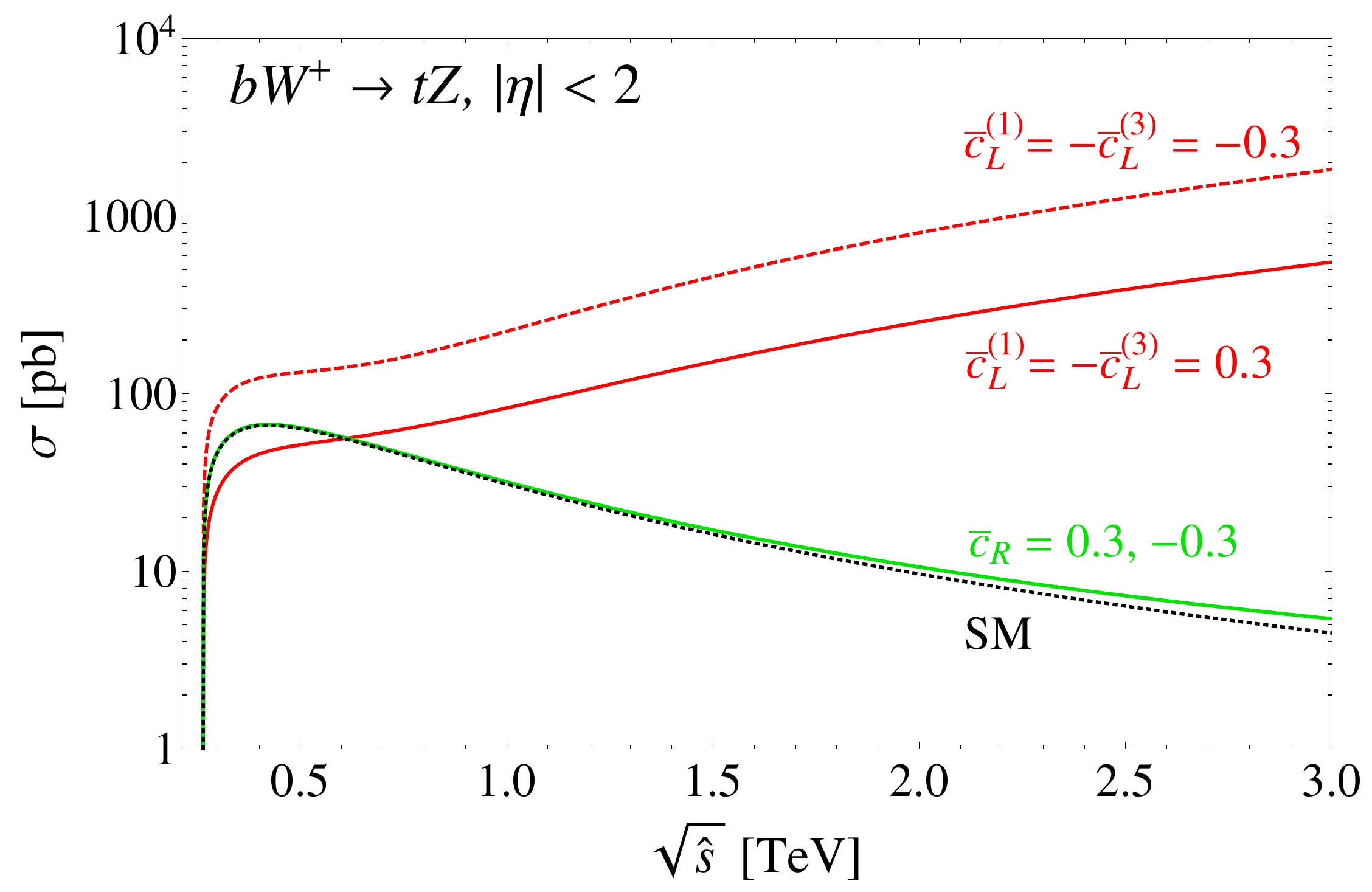}
 \end{center}
 \caption{Partonic cross section for the process $bW^{+}\to tZ$ as a function of the center of mass energy $\sqrt{\hat{s}}$. A pseudorapidity cut $\left|\eta\right|<2$ has been applied to remove the contribution of the forward region, which is enhanced by the diagram with $W$ exchange in the $t$-channel. At large energy, the red curves diverge like $\hat{s}$, the green curves (which are indistinguishable) tend to a constant limit, whereas the SM cross section (blue) falls off as $1/\hat{s}$.}
\label{fig:WbtZpartonic}
\end{figure}

For $ b W ^+ \rightarrow t Z $ we have $\kappa =1/\sqrt{2}$ and
\begin{align}
A_{LL} &\,=\,  \cLL \left[ 2-\cLt-\cLb + \tfrac{2}{3} s_w^2 (2 \cLt + \cLb - 3) \right], \nonumber \\
A_{RR} &\,=\,  \cRR \left[ 2 + \tfrac{2}{3} s_w^2 (2 \cRt + \cRb - 3) \right],\nonumber \\
A_{LR} &\,=\,  \cLL \left[ 1 - \cLb  + \tfrac{2}{3} s_w^2 (2 \cRt + \cLb - 3) \right],\nonumber\\
A_{RL} &\,=\,  \cRR \left[ 1 - \cLt + \tfrac{2}{3} s_w^2 (2 \cLt + \cRb - 3 ) \right].
\end{align}
This scattering can be probed at the LHC through $pp\to tZj$, which was already suggested in Refs.~\cite{tZ,RS} as a probe of the top-$Z$ couplings. For the pieces that grow like $\hat{s}$ we find $A_{LL}=2\cL(1-\cL)$ and $A_{RR}=0$, thus the leading sensitivity is to the coefficient $\cL$. In Fig.~\ref{fig:WbtZpartonic} we show the partonic cross section for $Wb\to tZ$ scattering. We observe that the cross section is significantly affected by a non-vanishing $\cL$, not only at large $\sqrt{\hat{s}}$ but also in the threshold region. On the contrary, $\cR$ has a very small impact on the cross section, because its effect arises via the subleading amplitude proportional to $A_{LR} = \cR (1-\cL)$, which grows only like $\sqrt{\hat{s}}$. In the SM, the cross section for $tZ$ production at the LHC is almost as large as the one for $t\bar{t}Z$, despite the fact that the former is a $b$-initiated pure electroweak process \cite{tZ}. This is due to the lower number of particles in the final state and the lower mass threshold. Notice that $tZj$ gives rise to the trilepton final state, therefore in principle it could be picked up by the CMS $8$ TeV $t\bar{t}Z$ search in the trilepton final state of Ref.~\cite{CMSttV}. However, the CMS event selection required at least four jets, among which at least two must be $b$-tagged, thus strongly suppressing the $tZj$ contribution. In fact, in Ref.~\cite{tZ} jet multiplicity was studied as a potential handle to distinguish $tZ$ from $t\bar{t}Z$ production. Based on these preliminary considerations, we conclude that $tZ$ production has negligible sensitivity to $\cR$, but may provide another opportunity to constrain the coefficient $\cL$.  
%

%
\subsection{$t Z \to t Z$}\label{tZ_tZ}
Next we consider $ t Z \rightarrow t Z $. We find $\kappa =1/2$ and
\begin{align}
A_{LL}&\,=\, A_{RR}=0\,,\nonumber \\
A_{LR}&\,=\,A_{RL}= (\cLt^2-\ct\cV) - \tfrac{8}{3} s_w^2 \cLt (\cLt - \cRt)+ \tfrac{16}{9}s_w^4 (\cLt - \cRt)^2\,.
\end{align}
This process can be probed in $pp\to t\bar{t}Zj$. Differently from $tW\to tW$, however, the $tZ\to tZ$ amplitude grows only linearly with energy, the corresponding coefficients $A_{LR}=A_{RL}$ depending on a combination of $\cL, \cR$ and $\cu$. As explained in Sec.~\ref{sec:CouplingsandEFT}, the absence of the $\hat{s}/v^2$ growth is a consequence of the symmetry structure of the $\chi \partial \chi \bar{\psi} \gamma \psi$ interactions. The sensitivity to $\cu$ is especially interesting, because it is absent in the dominant process for $t\bar{t}Z$ production, which is of $O(g_s^2 g_w)$ and only depends on $\cL,\cR$. Thus the $t\bar{t}Zj$ final state may in principle provide new information on $\cu$. The experimental strategy would rely on the trilepton final state, and the sensitivity to the $O(g_s g_w^3)$ contribution may be enhanced through a forward jet cut. Furthermore, since the amplitude grows with energy, a more stringent cut on the $ p _T $ of the $Z$ could also be effective.
\subsection{$t \bar{t} \to h h$}
The last process we consider is $ t \bar{t} \rightarrow h h $. The pieces of the amplitude that grow with energy can be written as\footnote{Here we define $\hat{s}=(p_t + p_{\bar{t}})^2$.}
\begin{equation} \label{amptthh}
\begin{pmatrix} \mathcal{M}_{LL} & \mathcal{M}_{RL} \\ \mathcal{M}_{LR} & \mathcal{M}_{RR} \end{pmatrix} =  -\frac{g^2}{m_W^2} \cth m_t\sqrt{\hat{s}}\begin{pmatrix} 0 & 1 \\
-1 & 0 \end{pmatrix}  \,\,,
\end{equation}
where to make the notation uniform with Eq.~\eqref{GeneralAmplitude}, in $\mathcal{M}_{ij}$ the index $i$ indicates the chirality of the top, and the index $j$ indicates the opposite of the chirality of the antitop. This process can be probed in $ p p \rightarrow t \bar{t}hh$, which was studies in detail in Ref.~\cite{Spannowskytthh}, with an emphasis on its potential role in constraining the Higgs cubic coupling. Based on the form of the amplitude in Eq.~\eqref{amptthh}, we observe that the main sensitivity of the cross section is not to deviations in the Higgs cubic coupling, which do not lead to a growth of the amplitude with energy, but to the $h^2t\bar{t}$ contact interaction (in turn proportional to $\bar{c}_u$), which leads to a growth like $m_t\sqrt{\hat s}/v^2$ of the amplitude. This conclusion is familiar from studies of the loop-induced process $gg\to hh$, which gives the largest contribution to the double Higgs production rate at the LHC and was shown to be strongly enhanced in presence of a $t\bar{t}hh$ interaction \cite{DoubleHiggstthh}. In fact, the $t\bar{t}\to hh$ amplitude can be obtained by performing an $s$-channel cut of $gg\to hh$. While $\cu$ will be constrained within $15\%$ by $t\bar{t}h$ production at LHC Run-2 \cite{SnowmassHiggs}, due to the growth of the amplitude with energy the residual effect in $pp\to t\bar{t}hh$ may be non-negligible, and potentially affect the Higgs cubic coupling constraint.    
 
\section{Conclusions}\label{sec:conclusions}
Progress towards an understanding of the weak scale requires testing the properties of the top quark. In natural models of electroweak symmetry breaking, the couplings of the latter generically deviate from their SM values. As long as the top couples strongly to the new physics resonances, such deformations can be large without requiring new light states. Examples of resonances are heavy vector bosons or vector-like fermions, typical of models where the Higgs arises from a strongly-interacting sector.

In this paper we proposed a new approach to measure deviations in the top electroweak couplings, which exploits the growth with energy of certain scattering amplitudes involving tops and longitudinal gauge bosons or Higgses. 
This high energy behaviour can be efficiently probed at the LHC, thanks to the large center of mass energies available.
As a proof of concept, we studied in detail $tW\to tW$ scattering, which diverges with the square of the energy in the presence of non-standard $ttZ$ couplings and can be studied at the LHC in $t\bar{t}Wj$ production.
By recasting an $8$ TeV CMS search for $t\bar{t}W$ in the same-sign lepton final state \cite{CMSttV}, we extracted constraints on the top-$Z$ couplings. We obtain improved limits compared to those derived from the ``conventional'' measurement of $t\bar{t}Z$ production, even though the analysis of Ref.~\cite{CMSttV} was not optimized for our signal. For example, considering only a deviation in the $Zt_R t_R$ coupling we find $-3.6 < \Delta_R < 2.4$ at $95\%$ CL. 

Having verified the effectiveness of our method, we proposed a dedicated $13$ TeV analysis. We exploited the distinctive kinematic properties of the $t\bar{t}Wj$ signal, namely a $tW$ pair with large invariant mass and a highly energetic forward jet, to suppress the background, mainly composed by ($t\bar{t}W$+jets)$_{\mathrm{QCD}}$ and misID$\ell$. Assuming $300$ fb$^{-1}$ of integrated luminosity and no systematic uncertainty on the background, we find $-0.83 < \Delta_R < 0.74$ at $95\%$ CL. In terms of the unique dim-6 operator that modifies the $Zt_R t_R$ coupling, this reads $-0.26 < \bar{c}_R < 0.23$. In the context of composite Higgs models with a fully composite $t_R$, where $\bar{c}_R \sim v^2/f^2$ with $f$ the Goldstone-Higgs decay constant, the bound translates into $f \gtrsim 500 \GeV$.

In addition, we identified several other amplitudes in the same class that could provide further evidence of the strong connection of the top quark with the new physics sector responsible for electroweak symmetry breaking. An interesting example is the $tZ\to th$ process, which is sensitive to modifications of $Zt_R t_R$ and can be probed at the LHC in $t\bar{t}hj$ production. It follows that $t\bar{t}h$+jets is sensitive to both of the two least known top couplings, namely $Zt_R t_R$ and $htt$, making it an ideal place to look for signs of BSM physics. This warrants further work, to fully exploit the opportunities offered by the LHC in testing the top-Higgs sector.

\section*{Acknowledgments}
We have benefited from discussions with F.~Maltoni. We thank B.~Mangano and A.~Brinkerhoff for correspondence about Refs.~\cite{CMSttV} and \cite{CMSmva}, respectively, and D.~Curtin and Y.~Tsai for discussions about the fake lepton simulation. JD and ES are very grateful to V.~Hirschi and O.~Mattelaer for their help with \mbox{MadGraph5\textunderscore aMC@NLO.} ES wishes to thank R.~R\"ontsch for correspondence about a typo in v1 of Ref.~\cite{RS}, which was eventually corrected in v2. The work of JD and MF is partly supported by the NSF through grant PHY-1316222. JD is supported in part by the NSERC Grant PGSD3-438393-2013, and MF by the DOE Grant DE-SC0003883. ES is supported by the DOE under Grant DE-SC-000999. JS has been supported in part by the MIUR-FIRB Grant RBFR12H1MW and the ERC Advanced Grant no.267985 (\emph{DaMeSyFla}). MF, ES and JS thank for hospitality and partial support (1) the Galileo Galilei Institute for Theoretical Physics, where this project was initiated, and (2) the Munich Institute for Astro- and Particle Physics of the DFG cluster of excellence ``Origin and Structure of the Universe,'' where part of the work was done.     

\appendix

\section{Electroweak Chiral Lagrangian} \label{App:chiL}
In the custodial invariant electroweak chiral Lagrangian the $SU(2)_L \times U(1)_Y$ SM gauge symmetry is non-linearly realized, with the Nambu-Goldstone bosons eaten by the $W$ and $Z$ parameterized by the $2 \times 2$ matrix
\beq 
\Sigma(x) = \exp \left(i \sigma^a \chi^a(x)/v \right) \ , 
\label{sigma}
\eeq
where $\sigma^a$ are the Pauli matrices. 
Such a $\Sigma$ field describes the spontaneous breaking $SU(2)_L \times SU(2)_R \to SU(2)_V$, with $U(1)_Y \subset SU(2)_R$.
The Higgs boson $h$ is introduced as a singlet under the custodial $SU(2)_V$ symmetry.

The interactions of the top (and bottom) are given, at the level of one derivative, by
\bea
\mathcal{L}^{\chi}_{t} \!\!\!&=&\!\!\! i \bar q_L \gamma^\mu D_\mu q_L + i \bar t_R \gamma^\mu D_\mu t_R + i \bar b_R \gamma^\mu D_\mu b_R \nonumber \\
&&\!\!\! - \ \frac{y_t v}{\sqrt{2}} \bar q_L \Sigma P_u t_R \left( 1 + \hat{c}_t \frac{h}{v} + 2 \hat{c}_t^h \frac{h^2}{v^2} + \cdots \right) + \mathrm{h.c.} \nonumber \\
&&\!\!\! - \ \frac{i}{2} \mathrm{Tr}\left[ \sigma^3 \Sigma^\dagger D_{\mu} \Sigma \right] \bar q_L \gamma^\mu q_L \left( \hat c_{L^{(1)}}  + 2 \hat c_{L^{(1)}}^{h} \frac{h}{v} + \cdots \right) \nonumber \\
&&\!\!\! + \ \frac{i}{2} \mathrm{Tr}\left[ \sigma^3 \Sigma^\dagger D_{\mu} \Sigma \right] \bar q_L \gamma^\mu \Sigma \sigma^3 \Sigma^\dagger q_L \left( \hat c_{L^{(2)}}  + 2 \hat c_{L^{(2)}}^{h} \frac{h}{v} + \cdots \right) \nonumber \\
&&\!\!\! + \ \frac{i}{2} \mathrm{Tr}\left[\Sigma^\dagger \sigma^a D_{\mu} \Sigma \right] \bar q_L \gamma^\mu \sigma^a q_L \left( \hat c_{L^{(3)}}  + 2 \hat c_{L^{(3)}}^{h} \frac{h}{v} + \cdots \right) \nonumber \\
&&\!\!\! - \ \frac{i}{2} \mathrm{Tr}\left[ \sigma^3 \Sigma^\dagger D_{\mu} \Sigma \right] \bar t_R \gamma^\mu t_R \left( \hat c_{R}  + 2 \hat c_{R}^{h} \frac{h}{v} + \cdots \right) \nonumber \\
&&\!\!\! - \ \frac{i}{2} \mathrm{Tr}\left[ \sigma^3 \Sigma^\dagger D_{\mu} \Sigma \right] \bar b_R \gamma^\mu b_R \left( \hat c_{R^{b}}  + 2 \hat c_{R^{b}}^{h} \frac{h}{v} + \cdots \right) \nonumber \\
&&\!\!\! + \ i P_u^T \Sigma^\dagger D_{\mu} \Sigma P_d \, \bar t_R \gamma^\mu b_R \left( \hat c_{R^{tb}}  + 2 \hat c_{R^{tb}}^{h} \frac{h}{v} + \cdots \right) + \mathrm{h.c.} \ ,
\label{Lchitop}
\eea
where the dots stand for higher order $h$ interactions. In \eq{Lchitop} we introduced $P_u = (1 , 0)^T$, $P_d = (0 , 1)^T$ as projectors onto the $Y = -1/2, +1/2$ components of $\Sigma$ respectively, and $D_\mu \Sigma = \partial_\mu \Sigma - ig W_\mu^a \sigma^a \Sigma/2 + i g' B_\mu \Sigma \sigma^3/2$. 
From \eq{sigma} one finds,
\bea
-\frac{i}{2} \mathrm{Tr}\left[ \sigma^3 \Sigma^\dagger D_{\mu} \Sigma \right] 
\!\!\!&\stackrel{\Sigma = 1}{=}&\!\!\! 
- \frac{g}{2c_w} Z_\mu \ , \nonumber \\
+ \frac{i}{2} \mathrm{Tr}\left[\Sigma^\dagger \sigma^a D_{\mu} \Sigma \right]
\!\!\!&\stackrel{\Sigma = 1}{=}&\!\!\! 
\frac{g}{2} W_{\mu}^{a} - \frac{g'}{2} B_{\mu} \delta^{a3} \ , \nonumber \\
i P_u^T \Sigma^\dagger D_{\mu} \Sigma P_d 
\!\!\!&\stackrel{\Sigma = 1}{=}&\!\!\! 
\frac{g}{\sqrt{2}} W_{\mu}^{+} \ ,
\label{devsigma}
\eea
in the unitary gauge $\Sigma = 1$, or equivalently at the leading order in the Nambu-Goldstone bosons $\chi^a$. 
The relations between the coefficients in \eq{Ltphen} and those in \eq{Lchitop} trivially follow,
\beq
\cLt = 1 +  \frac{- \hat c_{L^{(1)}} + \hat c_{L^{(2)}} + \hat c_{L^{(3)}}}{1-\frac{4}{3}s_w^2}
 \ , \quad 
\cLb = 1+ \frac{\hat c_{L^{(1)}} + \hat c_{L^{(2)}} + \hat c_{L^{(3)}}}{1-\frac{2}{3}s_w^2}
 \ , \quad 
\cLL = 1 + \hat c_{L^{(3)}}
 \ , \nonumber
\eeq
\beq
\cRt = 1 + \frac{\hat c_{R}}{\frac{4}{3}s_w^2}
 \ , \quad 
\cRb = 1 - \frac{\hat c_{R^{b}}}{\frac{2}{3}s_w^2}
 \ , \quad 
\cRR = \hat c_{R^{tb}} \ , \quad c_t = \hat{c}_t
 \ ,
\eeq
and similarly for the $\hat c_{i}^{h}$ coefficients.
Better suited to understand the high energy behaviour of scattering amplitudes is the gauge-less limit, $g, g' \to 0$. 
In that case one finds
\bea
-\frac{i}{2} \mathrm{Tr}\left[ \sigma^3 \Sigma^\dagger D_{\mu} \Sigma \right] 
\!\!\!&\stackrel{g,g' \to 0}{=}&\!\!\! 
\frac{1}{v} \partial_\mu \chi_3 + \frac{1}{v^2} \big( \chi_1 \partial_\mu \chi_2 - \chi_2 \partial_\mu \chi_1 \big) + O(\chi^3)
 \ , \nonumber \\
+ \frac{i}{2} \mathrm{Tr}\left[\Sigma^\dagger \sigma^a D_{\mu} \Sigma \right]
\!\!\!&\stackrel{g,g' \to 0}{=}&\!\!\! 
- \frac{1}{v} \partial_\mu \chi_a + \frac{1}{v^2} \epsilon_{abc} \chi_b \partial_\mu \chi_c + O(\chi^3)
 \ , \nonumber \\
i P_u^T \Sigma^\dagger D_{\mu} \Sigma P_d 
\!\!\!&\stackrel{g,g' \to 0}{=}&\!\!\! 
- \frac{\sqrt{2}}{v} \partial_\mu \chi_+ + i \frac{\sqrt{2}}{v^2} \big( \chi_3 \partial_\mu \chi_+ - \chi_+ \partial_\mu \chi_3 \big) + O(\chi^3) \ .
\label{devsigmagaugeless}
\eea

In a similar fashion one can write the leading interactions of the Higgs boson, at the level of two derivatives,
\bea
\mathcal{L}^{\chi}_{h} \!\!\!&=&\!\!\! \frac{1}{2} (\partial_\mu h)^2 
+ \frac{v^2}{4} \mathrm{Tr} [|D_\mu \Sigma|^2] \left( 1  + 2 \hat{c}_V \frac{h}{v} + \cdots \right) 
\nonumber \\
&&\!\!\! - \ \frac{1}{2} m_h^2 h^2 - \hat{c}_3 \frac{m_h^2}{2v} h^3 + \cdots \ ,
\label{Lchihiggs}
\eea
where the dots stand for higher order $h$ interactions. The relation to Eq.~\eqref{Lhphen} is given by $\cV = \hat{c}_V$ and $\ch = \hat{c}_3$.

\section{Current and projected $t\bar{t}Z$ constraints} \label{App:ttZ}
Here we discuss briefly the constraints derived from the $t\bar{t}Z$ process, both using $8$ TeV data \cite{CMSttV} and an existing projection to $13$ TeV \cite{RS}, which we used for comparison with our bounds obtained from $t\bar{t}W$. 
\subsection{8 TeV $t\bar{t}Z$ bound}
The trilepton analysis in Ref.~\cite{CMSttV} was targeted at measuring the $t\bar{t}Z$ process, and thus requires, in addition to two of the leptons being compatible with a $Z$ decay, at least $4$ jets, among which at least $2$ are $b$-tagged. To set a limit on the parameters $\vec{p}$ from that analysis, we make use of the event yields listed in Table~2 of Ref.~\cite{CMSttV} and we assume a systematic uncertainty of $50\%$ on the total background\footnote{We have verified that by assuming $50\%$ on the total background as the only systematic uncertainty, we reproduce to good accuracy the measurement of the $t\bar{t}Z$ cross section quoted in Ref.~\cite{CMSttV}: we find $197^{+107}_{-97}$ fb, to be compared with $190^{+108}_{-89}$ fb.}
\begin{equation}
L(\vec{p}\,;r) = \frac{(N_{S+B})^{N_{obs}}e^{-N_{S+B}}}{N_{obs}!} P_{0.5}(r,1)
\end{equation}
where $P_{\sigma}(x,x_0)$ was defined in Eq.~\eqref{likelihood}, and $N_{S+B}= r N_B + \sigma_{t\bar{t}Z} (\vec{p}) \mathcal{L}\epsilon$, with $\sigma_{t\bar{t}Z}$ the inclusive cross section for $pp\to t\bar{t}Z$ at $8$ TeV, $\mathcal{L}=19.5\;\mathrm{fb}^{-1}$ the integrated luminosity and $\epsilon$ the total efficiency for the SM $t\bar{t}Z$ process. The assumption of constant efficiency is justified, given that the cross section does not grow with energy for non-SM couplings. CMS finds that the contribution of $(t\bar{t}W$+jets$)_{\mathrm{QCD}}$ to the signal region is strongly subleading, therefore the sensitivity to the couplings arising from $(t\bar{t}Wj)_{\mathrm{EW}}$ is negligible.

\subsection{13 TeV $t\bar{t}Z$ projection}
The most recent assessment of the projected LHC sensitivity to top-$Z$ couplings in the $pp\to t\bar{t}Z$ process was performed in Ref.~\cite{RS}, by making use of a signal computation at NLO in QCD. The authors focused on the trilepton final state, and to set constraints they exploited, in addition to the total cross section, the differential distribution in the azimuthal opening angle between the leptons stemming from the $Z$ decay. Neither backgrounds nor detector effects were considered. To compare with our results we make use of their Fig.~10, where the relation $\bar{c}_L^{\,(1)} + \bar{c}_L^{\,(3)} =0$ was assumed, and simply map the exclusion contours given there to the planes $(\Delta_L, \Delta_R)$ and $(\cL, \cR)$ used in this paper.    

\section{Fake lepton simulation} \label{App:FakeLeptons}
We follow Ref.~\cite{FakeLeptons}, which proposed a method to efficiently simulate fake leptons starting from MC samples containing jets. The method exploits the relationship between the kinematics of a fake lepton and that of the jet that `sources' it. It consists in applying to each jet an \emph{efficiency} to generate a fake lepton, assumed to be a function of the jet $p_T$, and a \emph{transfer function}, which represents a normalized probability distribution for the fraction of the jet $p_T$ that is inherited by the fake lepton. These are parameterized as follows
\begin{align}
\epsilon_{j\to \ell}(p_T^j) \,=&\,\, \epsilon_{200}\left[1-(1-r_{10})\frac{200-p_T^j/\mathrm{GeV}}{200-10}\right]\,, \\
\mathcal{T}_{j\to \ell}(\alpha) \,=&\, \left(\frac{\sqrt{2\pi}\sigma}{2}\right)^{-1} \left[\mathrm{erf}\left(\frac{1-\mu}{\sqrt{2}\sigma}\right)+\mathrm{erf}\left(\frac{\mu}{\sqrt{2}\sigma}\right)\right]^{-1}e^{-\frac{(\alpha-\mu)^2}{2\sigma^2}},
\end{align}
where $\alpha\equiv 1-p_T^\ell/p_T^j\,$ is the fraction of the jet momentum that is not transferred to the fake lepton. The residual momentum is assumed to contribute to the MET. The parameter $\epsilon_{200}$ represents the efficiency for fake lepton production at $p_T^j = 200\;\mathrm{GeV}$, whereas $r_{10}$ sets the slope of the efficiency as function of $p_T^j$. The transfer function is assumed to be a Gaussian with mean $\mu$ and standard deviation $\sigma$. In our analysis, the `source' process is $t\bar{t}+\mathrm{jets}$, and we will assume that fake leptons dominantly originate from heavy flavor ($b$) jets \cite{CMSttV}. The parameters of the fake lepton simulation are chosen as follows. We first set, for simplicity, $r_{10}=1$, which gives an efficiency independent of the jet $p_T$. We further set $\mu=0.5$, based on the generic expectation of equal splitting of the momentum between the fake lepton and the neutrino produced in heavy flavor decays. By comparison with the $H_T$ and $p_T^{\ell_1}$ distributions by CMS, which were obtained with a data-driven method and reported in Fig.~2 of Ref.~\cite{CMSttV}, we find that $\sigma = 0.1$ gives reasonable agreement. We are thus left with only one free parameter, the global efficiency, which we fix to $\epsilon_{200}\approx 2.5\times 10^{-4}$ to  reproduce the total event yield of $12.1$ quoted by CMS (see Table~\ref{tab:8TeVyields}). A somewhat similar choice of parameters was made by the authors of Ref.~\cite{ShuveEtal}. We assume no significant difference occurs between fake lepton production at $8$ and $13$ TeV, and employ the above values of the parameters in our $13$ TeV analysis.

\end{document}